\documentclass[journal,comsoc]{IEEEtran}
\usepackage{amsfonts}
\usepackage{amssymb}
\usepackage{eurosym}
\usepackage{cite}
\usepackage{graphicx}
\usepackage{epstopdf}
\usepackage{amsmath}
\usepackage{tikz,lipsum}
\usepackage{caption}
\usepackage[T1]{fontenc}
\usepackage{amsthm}
\usepackage{mathrsfs}
\usepackage{color}
\usepackage{stfloats}
\usepackage{xcolor, etoolbox}
\usepackage{subcaption}
\usepackage{comment}
\usepackage{algcompatible}
\usepackage[ruled,linesnumbered]{algorithm2e}
\usepackage{bm}
\usepackage{upgreek}
 \def\baselinestretch{.99}
\IEEEoverridecommandlockouts
\newcounter{mytempeqncnt}
\newtheorem{remark}{Remark}
\newtheorem{theorem}{Theorem}

\begin{document}

\title{Beyond-Diagonal RIS For Enhanced Secrecy and Sensing Gains in Secure
ISAC Networks: An Optimization Framework}
\author{Elmehdi Illi, \IEEEmembership{Member, IEEE}, and Marwa Qaraqe, 
\IEEEmembership{Senior
Member, IEEE} \thanks{%
E. Illi and M. Qaraqe are with the College of Science and Engineering, Hamad
Bin Khalifa University, Doha, Qatar. (e-mails: elmehdi.illi@ieee.org,
mqaraqe@hbku.edu.qa.)}}
\maketitle

\begin{abstract}
Integrated sensing and communication (ISAC) has been receiving a notable
interest as an energy- and spectrum-efficient enabler for simultaneous
communication and sensing. Notably, reconfigurable intelligent surfaces
(RIS) is among the key technologies enabling robust communication and
sensing, particularly in environments without a line-of-sight (LoS).
Recently, a new type of RIS, called beyond-diagonal RIS (BD-RIS), has drawn
attention, offering additional degrees of freedom in controlling the
propagation medium. In this paper, a novel secure BD-RIS-aided ISAC scheme
is proposed and evaluated. The scheme is applicable to a multi-user
multi-target ISAC network, where a dual-functional radar-communication
(DFRC) base station (BS) simultaneously serves multiple downlink users and
senses various targets that aim to eavesdrop on the legitimate signal
transmitted to the users. The presence of a BD-RIS enables circumventing the
absence of the LoS link and ensures secure transmission and sensing. To this
end, an optimization problem is formulated aiming at maximizing a weighted
sum of per-target reflected powers, subject to secrecy and transmit power
constraints. Thus, by virtue of an Augmented Lagrangian- and Riemannian conjugate gradient-based approach, in addition to semidefinite programming, an alternating optimization (AO)-based algorithm is developed, which provides a local optimum for the BD-RIS scattering matrix, transmit signal beamforming matrices, and artificial noise covariance matrix. Numerical results highlight (i) the notable sensing gains of the BD-RIS-aided design with respect to its
diagonal RIS (D-RIS)-based baseline and (ii) the improved secrecy-sensing
trade-off, whereby the BD-RIS can ensure an increasing system secrecy
without a significant loss in the per-target reflected power.
\end{abstract}

\begin{IEEEkeywords}
Dual-functional radar communication (DFRC), eavesdropping, integrated sensing and communication (ISAC), physical-layer security, and beyond-diagonal reconfigurable intelligent surfaces (BD-RIS).
\end{IEEEkeywords}

\section{Introduction}

\subsection{Background}

Integrated sensing and communication (ISAC) technology has been gaining
increased interest as an enabling solution for establishing a spectrum- and
power-efficient sensing and communication tasks. ISAC is based on utilizing
the same frequency spectrum, power, and hardware modules to perform both
communication and sensing tasks using the same signal. From a deployment
perspective, dual-function radar communication (DFRC) stands out as one of
the enabling approaches, whereby a communication signal is embedded in radar
waveform or vice-versa \cite{dfrc}. DFRC-based ISAC schemes are based on
transmitting both communication and radar beams from a DFRC unit, e.g., base
station (BS), by leveraging a multiple-antenna transmit array. The same unit
collects back the reflected echo by the target using another array of
receive antennas collocated with the transmit ones.

The lack of line-of-sight (LoS) remains a major challenge in wireless
communication and RF sensing, as multiple-input multiple-output (MIMO)
technology is unable to turn non-LoS channels into LoS ones. On another
front, ISAC waveforms introduce a data confidentiality risk, since sensed
targets could act as eavesdroppers aiming at illegitimately accessing
confidential data. Such a threat is expected to be amplified with the
evolution of quantum computing technology. Physical-layer security (PLS)
addresses this by using stochastic encoding and transmission techniques
(e.g., MIMO precoding, receive diversity) to ensure information-theoretic
secure communication when the legitimate channel is stronger than the
eavesdropper's. Designing ISAC systems that accurately sense malicious
targets while maintaining secure communication is thus crucial.

In response to the aforementioned challenges, reconfigurable intelligent
surfaces (RIS) technology has emerged as a rapidly evolving solution capable
of adapting the wireless channel into a favorable one, avoid the LoS
obstruction/absence issue, and counter eavesdropping. An RIS is a planar
metasurface with several reflecting elements (REs), often called meta-atoms,
each with a tunable impedance controlling its reflection coefficient. Thus,
an incoming signal to each RE can be adjusted in phase and amplitude to
control its reflection direction, which can improve the received signal
level in the locations of interest, e.g., users' or targets' zone, and/or
decrease it in undesired areas, e.g., malicious targets, improving ISAC
networks' reliability, security, and detection accuracy \cite{lit1}.
However, despite the above potential of RIS, its gains are limited by its
reflection coefficients matrix. The latter is a diagonal matrix whose
elements correspond to the reflection coefficient (phase shift) for the
various REs, while the off-diagonal elements are set to zero. Inspired by
the potential generalization of the RIS diagonal phase shift matrix into a
non-diagonal one, a new RIS variant, called beyond-diagonal (BD)-RIS, was
proposed. A BD-RIS offers a more flexible design by exploiting all the
elements of the reflection matrix, thereby enhancing the beamforming gain
and providing additional degrees of freedom for signal beam steering \cite%
{tutorialbdris}. From a design perspective, such an enhanced scheme
translates into creating electronic interconnections between the different
REs using adjustable impedances to allow incoming signal waves to each RE to
be partly reflected by other elements, producing non-zero entries in the
reflection matrix.

\subsection{Related Work}

Despite its recent introduction compared to the traditional diagonal RIS
(D-RIS), a considerable amount of work in the literature is noticed. Some of
the research work investigated different BD-RIS architectures with the aim
of maximizing the received signal power in different network setups, such as
in multiple-input single-output (MISO) and MIMO setups with fully- and
group-connected architectures \cite{lit1,lit2,lit3,lit4,lit5}. 
From a communication rate or sum rate maximization perspective, the work in 
\cite{lit6} proposed a closed-form solution for the BD-RIS scattering matrix
maximizing the sum rate of a single-input single-output network. The authors
of \cite{lit7} tackled the theoretical rate maximization of a point-to-point
BD-RIS-aided MIMO system. The authors analyzed the shaping potential of
BD-RIS in MIMO networks by inspecting its singular values. Furthermore, a
numerical optimization framework for optimizing the BD-RIS scattering matrix
was proposed. Another similar analysis was carried out in \cite{lit8}, where
a MIMO system's capacity was maximized by optimizing the BD-RIS scattering
matrix and the transmit covariance matrix. In \cite{lit10}, the authors
proposed a BD-RIS scattering matrix design maximizing MIMO systems'
capacity, by formulating the unitary scattering matrix as a product of two
unitary ones, aligned to the forward and backward channel matrices. From a
multi-user perspective, the work in \cite{lit11} proposed an alternating
optimization (AO)-based low-complexity closed-form solution for the BD-RIS
scattering matrix, aiming at maximizing the sum channel gains. The authors
of \cite{lit12} proposed a two-stage fractional programming-based approach
to optimize a reciprocal BD-RIS scattering matrix and the BS's beamforming
matrix. The authors of \cite{lit14} proposed a polygonal multi-sector BD-RIS
design, comprising a RIS on each facade of the polygonal prism
interconnected with other facades' RISs. A representative model for the
propagation channel was proposed, along with a scaling law analysis and
BD-RIS configuration optimization to maximize the network's sum rate. The
same authors proposed in \cite{lit13} a dynamic grouping strategy for a
group-connected BD-RIS design, enabling a robust adapting to the channel
state information (CSI) of the different users. Samy et al. conducted in 
\cite{lit15} a thorough analysis of multi-sector BD-RIS designs and compared
its reliability performance with that of the simultaneous transmitting and
reflecting RIS (STAR-RIS), representing a particular-case design of the
multi-sector one.

As far as wireless networks' PLS is concerned, only a few works were
reported to analyze the impact of BD-RIS on the wireless network secrecy.
The work in \cite{lit16,lit17} proposed BD-RIS-assisted secure transmission
schemes, serving a multi-antenna legitimate user, in the presence of a
multi-antenna eavesdropper. A projected gradient-based approach was proposed
to maximize the system's secrecy capacity (SC). Furthermore, in \cite{lit19}, the ergodic SC of a single-user single-eavesdropper BD-RIS-aided system
was analyzed. Also, the authors of \cite{lit18} analyzed the secrecy
performance of a BD-RIS multi-user network in the presence of an aerial
eavesdropper. A user scheduling scheme was proposed to serve a single user
at once by the BD-RIS, enabling a simple closed-form solution for its
reflection matrix. In \cite{lit20}, a BD-RIS-assisted cognitive
radio-enabled non-terrestrial network was analyzed, in the presence of a
malicious ground eavesdropper. A projected gradient-based scheme, maximizing
the system's SC, was proposed.

On another front, it is worth depicting several research works that analyzed
the reliability and sensing gains of different BD-RIS-assisted ISAC
networks. For instance, the authors of \cite{lit21} proposed a transmissive
and reflective BD-RIS design covering a wider space of potential targets and
users. The proposed scheme focused on optimizing the BD-RIS reflection
coefficients, maximizing the sensing signal-to-noise ratio (SNR) subject to
communication rate constraints. A similar setup was analyzed in \cite{lit22}
by swapping the objective and constraint functions. In \cite{lit24}, an
optimization of the BD-RIS reflection matrix was carried out by maximizing a
linear combination between sensing and communication SNRs. Also, the authors
of \cite{lit23} developed an optimization framework for minimizing the
network's power subject to communication rate and sensing constraints. An
analysis and maximization of the reliability performance of an ISAC network,
subject to sensing constraints, was tackled in \cite{lit26}. Importantly,
the proposed scheme considered an unknown target location with only
statistical information about it.

\subsection{Motivation and Contributions}

In spite of the above-mentioned research efforts on assessing and optimizing
BD-RIS-aided networks, with the objective of improving reliability, sensing,
secrecy, or power consumption, it should be noted that no prior analysis on
BD-RIS was conducted on secure ISAC networks. In such communication systems,
sensed targets can act as potential eavesdroppers. Therefore, it is crucial
to maintain an information-theoretic data confidentiality, e.g., no
legitimate data leakage to the targets, in addition to preserving
reliability and sensing requirements. Motivated by the above, in this work,
an enhanced BD-RIS-aided secure ISAC scheme is proposed and analyzed. The
proposed scheme is applicable to a multi-user multi-target ISAC network,
where a fully-connected non-reciprocal BD-RIS is involved to enable signal
reflection. An AO-based framework is proposed, based on Riemannian conjugate gradient (RCG) for optimizing the the BD-RIS scattering matrix, whereas transmit beamforming, and AN covariance matrices are optimized via semidefinite relaxation (SDR) to maximize the sensing performance, subject to secrecy and transmit power constraints. The proposed BD-RIS scheme demonstrates notable gains compared to a D-RIS baseline one in terms of per-target and sum reflected power. The main contributions of this paper are summarized as follows:

\begin{itemize}
\item A BD-RIS-aided secure ISAC network, comprising several downlink
legitimate users and malicious targets, is studied and modeled. In
particular, the main secrecy and sensing metrics are formulated in terms of
the key system parameters, such as the BD-RIS scattering matrix, transmit
beamforming matrices, and the AN covariance matrix. 

\item An AO-based optimization framework is proposed, based on decomposing the
optimization for the (i) BD-RIS scattering matrix and the (ii) transmit
signal and AN covariance matrices into two subproblems, sequentially solved. 

\item To tackle the non-convex BD-RIS scattering matrix optimization
subproblem, an RCG approach is utilized, which is based on (i) building a penalty-based Augmented Lagrangian (AL) from the primal subproblem, (ii) performing a manifold optimization by virtue of gradient descent rules on the Stiefel manifold, and (iii) updating the AL's penalty coefficient multipliers. To enhance solution quality, the unitary BD-RIS scattering matrix is decomposed into the product of an initial suboptimal BD-RIS matrix solution and an iteratively optimized unitary matrix.

\item A proof of the convergence of the aforementioned proposed iterative framework is provided, demonstrating the convergence to a local optimal solution that fulfills the Karush-Kuhn Tucker (KKT) conditions.

\item Comprehensive simulations are performed to evaluate the proposed
scheme's secrecy and sensing performance. Notably, the obtained results
demonstrate a convergence of the proposed AO-based approach within a finite number of iterations, in addition to the fulfillment of the preset
secrecy constraints.
\end{itemize}

The remainder of this paper is organized as follows: Section II is dedicated
to present the adopted network and channel models along with communication,
eavesdropping, and sensing assumptions and metrics. Section III presents the
tackled optimization problem and details the adopted solution's algorithmic approach.
In Section IV, illustrative numerical results are provided to evaluate the
secrecy and sensing performance of the proposed scheme. Finally, Section V
concludes the paper.

\subsection{Notations}

$(.)^T$ and $(.)^H $ denote, respectively, the transpose and the conjugate
transpose (Hermitian) of a vector or matrix, $\otimes $ is the Kronecker
product, $\left[ \mathbf{Z}\right] _{m,:}$ and $\left[ \mathbf{Z}\right]
_{:,n}$ denote, respectively, the $m$th row and $n$th column of $\mathbf{Z}$%
, $\left[ \mathbf{Z}\right] _{m,n}$ is the element at the $m$th row and $n$%
th column of $\mathbf{Z}$, $\mathbf{I}_{M}$ is the $M\times M$ identity
matrix, $\mathbb{E}\left[ \mathbf{.}\right] $ is the expectation operator,
and $\mathrm{Tr}\left[ \mathbf{.}\right] $ is the trace of a matrix. Additionally, $%
\left\lfloor .\right\rfloor $ is the floor of a real value. Also, $\mathrm{St%
}\left( M,M\right) $ denotes the Stiefel manifold, defined as $\mathrm{St}%
\left( M,M\right) =\left\{ \mathbf{U\in 
\mathbb{C}
}^{M\times M}\left\vert \mathbf{U}^{H}\mathbf{U=I}_{M}\right. \right\} $, $%
\nabla _{\mathrm{E,}\mathbf{x}}\mathcal{F}\left( \mathbf{x}\right) $ denotes
the Euclidean gradient on $\mathbf{x}$ of a function $\mathcal{F}$, while $%
\nabla _{\mathrm{R,}\mathbf{X}}\mathcal{F}\left( \mathbf{x}\right) $ is its
Riemannian gradient computed on the Stiefel manifold, and $\mathrm{trans}
_{Q\leftarrow P}\left( .\right) $ defines a transportation operation of a
vector/matrix from a tangent space at a point $P$ to the tangent space at
another point $Q$. Lastly, $\mathrm{retr}\left( \mathbf{X}\right) $ defines
a retraction operation for a point $\mathbf{X}\in \mathbb{C}^{M\times M}$
mapping it to the closest point $\tilde{\mathbf{X}} \in \mathrm{St}(M,M)$.

\section{System and Channel Model}

\begin{figure}[h]
\vspace*{-.5cm}
\par
\begin{center}
\includegraphics[scale=.2]{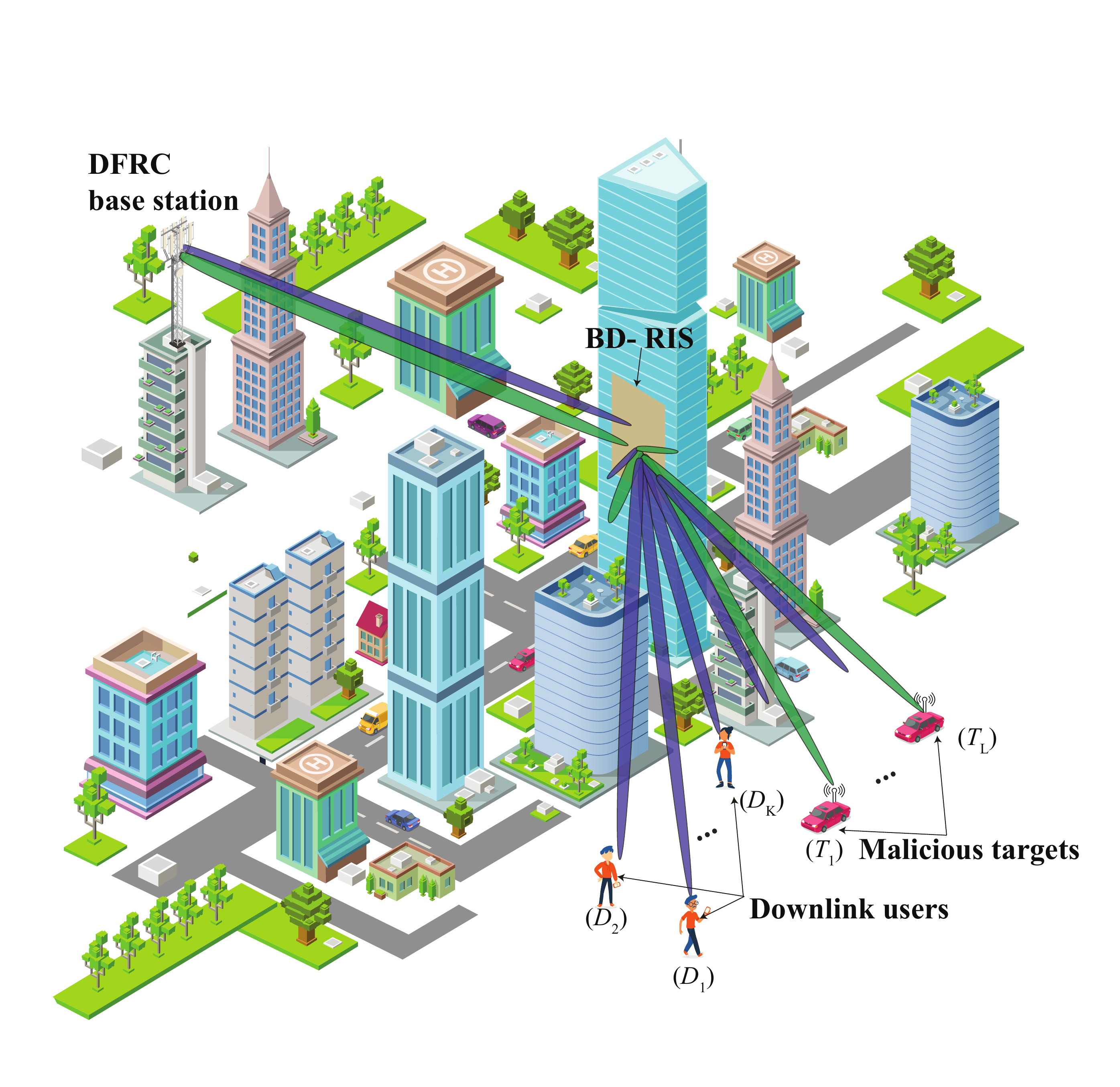}
\end{center}
\par
\vspace*{-.2cm}
\caption{Considered System model.}
\label{figsys}
\end{figure}
\subsection{Network Model}
Let us consider a secure ISAC wireless network given in Fig. \ref{figsys},
consisting of a dual-functional radar communication (DFRC) base station
(BS), denoted by $S$, serving $K$ downlink users $\{D_{k}\}_{k=1}^{K}$ while
simultaneously sensing $L$ targets $\{T_{l}\}_{l=1}^{L}$. $S$ aims at
performing a target detection to confirm the presence of the targets in the
pre-known locations. Additionally, the set of $L$ targets are assumed to be
malicious entities aiming at compromising the communication confidentiality
by eavesdropping on the legitimate communication signals of the $K$ users.
Thus, $S$ broadcast via an array of $J_{\mathrm{T}}$ transmit antenna an
ISAC\ signal designed to comprise both the legitimate information signal
directed towards the users as well as an artificial noise (AN) dedicated to
sense the presence of the targets and preserve communication
confidentiality. Furthermore, $S$ is equipped by an additional array of $J_{%
\mathrm{R}}$ receive antennas to process the echo signals reflected back from
the targets. Due to the possible presence of obstacles in the considered
environment, direct-link transmissions tend to be obstructed. Thus, an $M$%
-REs RIS\ $(R)$ with a beyond-diagonal reflection pattern\ is incorporated
to assist the ISAC\ signal propagation by creating a virtual LoS link
between $S$ and the different users and targets. As a result, one can
express the received downlink signal at the users as well as the malicious
eavesdroppers as 
\begin{equation}
y_{\Omega}=\mathbf{h}_{S\Omega }\mathbf{v}+n_{\Omega },\Omega \in \left\{
D_{k},T_{l}\right\} .  \label{rxsig1}
\end{equation}%
In \eqref{rxsig1}, 
\begin{equation}
\mathbf{h}_{S\Omega }\triangleq \sqrt{\zeta _{S\Omega }}\mathbf{h}_{R\Omega }%
\mathbf{\Psi H}_{SR}\in \mathbb{C}^{1\times J_{\mathrm{T}}}  \label{hsomega}
\end{equation}%
represents the cascaded channel vector of the link between $S$ and $\Omega$
via $R $ with 
\begin{equation}
\zeta _{S\Omega }=\frac{\mathcal{G}_{S,\mathrm{T}}\mathcal{G}_{\Omega ,%
\mathrm{R}}\rho _{R}^{4}}{d_{SR}^{2}d_{R\Omega }^{2}\left( 4\pi \right) ^{2}}
\label{fsplris}
\end{equation}%
defining the corresponding link's cascaded free-space path loss (FSPL),
where $\mathcal{G}_{S,\mathrm{T}}$ and $\mathcal{G}_{\Omega ,\mathrm{R}}$
are the antenna gain of $S$'s transmit antennas\footnote{%
For the sake of simplicity, we assume that the transmit antennas at the BS
are of the same transmit gain.} and $\Omega $'s receive one, respectively.
Additionally, $\rho_R $ is the size of each element of $R$, and $d_{SR}$ and 
$d_{R\Omega }$ denote the euclidean distances of the $S$-$R$ and $R$-$\Omega$
links, respectively. Furthermore, the channel matrix $\mathbf{H}_{SR}\in 
\mathbb{C}^{M\times J_{\mathrm{T}}\text{ }}$and the channel vector $\mathbf{h%
}_{R\Omega}\in \mathbb{C}^{1\times M}$ comprise the small-scale fading
coefficients corresponding to the $S$-$R$ and $R$-$\Omega $ channels, which
can be expressed as 
\begin{equation}
\mathbf{H}_{SR}=\sqrt{\frac{\mathcal{K}_{SR}}{\mathcal{K}_{SR}+1}}\mathbf{H}%
_{SR}^{\mathrm{(L)}}+\sqrt{\frac{1}{\mathcal{K}_{SR}+1}}\mathbf{H}_{SR}^{%
\mathrm{(NL)}}  \label{Hbr}
\end{equation}%
and 
\begin{equation}
\mathbf{h}_{R\Omega }=\sqrt{\frac{\mathcal{K}_{R\Omega }}{\mathcal{K}%
_{R\Omega }+1}}\mathbf{h}_{R\Omega }^{\mathrm{(L)}}+\sqrt{\frac{1}{\mathcal{K%
}_{R\Omega }+1}}\mathbf{h}_{R\Omega }^{\mathrm{(NL)}} . \label{hrz}
\end{equation}%
Observe from \eqref{Hbr} and \eqref{hrz} that both $\mathbf{H}_{SR}$ and $%
\mathbf{h}_{R\Omega}$ and are composed of a deterministic LoS term as well
as a non-LoS (NLoS) one. On the one hand, the former, e.g., the array in the
first term of \eqref{Hbr} and \eqref{hrz}, can be expressed as:%
\begin{equation}
\mathbf{H}_{SR}^{\mathrm{(L)}}=\mathbf{e}^{T}\left( \theta _{SR},\phi
_{SR}\right) \mathbf{f}\left( \theta _{SR},\phi _{SR},\rho _{S},J_{\mathrm{T}%
}\right)  \label{Hbrlos}
\end{equation}%
and 
\begin{equation}
\mathbf{h}_{R\Omega }^{\mathrm{(L)}}=\mathbf{e}\left( \theta _{R\Omega},\phi
_{R\Omega}\right) \in \mathbb{C}^{1\times M} , \label{hrzlos}
\end{equation}%
where 
\begin{equation}
\mathbf{e}\left( \theta ,\phi \right) =\mathbf{f}\left( 0,\theta ,\rho
_{R},M_{a}\right) \otimes \mathbf{f}\left( \theta ,\phi ,\rho
_{R},M_{b}\right),  \label{e}
\end{equation}%
\begin{equation}
\mathbf{f}\left( \theta ,\phi ,\beta ,N\right) =\left[ 1,\ldots ,e^{-j2\pi
\beta /\lambda \left( N-1\right) \cos \theta \sin \phi }\right] \in \mathbb{C%
}^{1\times N},  \label{f}
\end{equation}%
$M_{a}$ and $M_{b}$ represent the number of $R$'s elements
horizontally and vertically, respectively, with $M=M_{a}M_{b}$. Also, $%
\theta _{VW}$ and $\phi _{VW}$ $\left( \forall VW\in \left\{
SR,RD_{k},RT_{l}\right\} \right) $ are, respectively, the relative elevation
and azimuth angles between a pair of nodes $V$ and $W$ with respect to the
broadside direction of either $S$'s transmit array or $R$. In addition, $%
\rho _{S}$ represents the inter-antenna spacing at the transmit array of $S$%
, while $\lambda $ denotes the signal wavelength, and $\mathcal{K}_{VW}$ is
the Rician $K$-factor of the $V$-$W$ link. On the other hand, the NLoS\
terms of the $S$-$R$ and $R$-$\Omega$ link, i.e., $\mathbf{H}_{SR}^{\mathrm{%
(NL)}}$ and $\mathbf{h}_{R\Omega}^{\mathrm{(NL)}}$ are composed of zero-mean
complex Gaussian-distributed elements with covariance matrices $\mathbb{E}[[%
\mathbf{H}_{SR}^{\mathrm{(NL)}}]_{:,m}([\mathbf{H}_{SR}^{\mathrm{(NL)}%
}]_{:,m})^{H}]=\mathbb{E}[(\mathbf{h}_{R\Omega }^{\mathrm{(NL)}})^{H}\mathbf{%
h}_{R\Omega }^{\mathrm{(NL)}}]=\mathbf{I}_{M}$ $(\forall m)$.

Furthermore, $\mathbf{\Psi }$ is defined as the BD-RIS\ reflection matrix.
Such a particular RIS\ structure is enabled by interconnecting the REs of
each RIS\ via tunable impedance-based circuits, which allows each element to
transfer a portion of the signal power reaching it to other REs in the RIS 
\cite{tutorialbdris}. This allows every RE to reflect a mixture of the
signal impinging on it and signals reaching other REs. Accordingly, several
BD-RIS\ architectures and topologies can be established as a function of the
number of interconnections. For instance, a fully-connected (FC) BD-RIS
network consists of an array of REs where each element is connected to all
the other $M-1$ ones, in addition to another connection to ground via
another tunable impedance. Also, the BD-RIS design can exhibit symmetry if
the same impedance circuitry element links a pair of REs in both directions,
i.e., $\mathbf{\Psi}^T=\mathbf{\Psi}$. The generalizing architecture is an
asymmetric BD-RIS design consisting of distinct circuit elements that link
each pair of REs in the two directions. Thus, an FC-based asymmetric BD-RIS\
architecture is considered in this work, whereby $R$'s reflection matrix can
be expressed as%
\begin{equation}
\mathbf{\Psi =}\left[ 
\begin{array}{ccc}
\psi _{1,1} & \ldots & \psi _{1,M} \\ 
\vdots & \ddots & \vdots \\ 
\psi _{M,1} & \ldots & \psi _{M,M}%
\end{array}%
\right] \in \mathbb{C}^{M\times M},  \label{Psibd}
\end{equation}%
where $\psi _{m,n}$ is the complex-valued reflection coefficient describing
the reflection gain by the $m$th RE $(\mathrm{R}_{m})$ of the signal
reaching $\mathrm{R}_n$. Unlike the D-RIS, observe that the BD-RIS'
reflection matrix in \eqref{Psibd} extends the diagonal one by also allowing
off-diagonal elements to be non-zero, which offers extra degrees of freedom
to control signal beamsteering as well as to increase the received signal
power at the intended user and/or target. In this optic, $\mathbf{\Psi }$ is
defined as a unitary matrix fulfilling the following property:\ $\mathbf{%
\Psi }^{H}\mathbf{\Psi =I}_{M}$. Moreover, we define $\mathbf{v}=\mathbf{Qx}+%
\mathbf{z}$ as the ISAC signal transmitted by $S $ where $\mathbf{Q}%
\triangleq \left[ \mathbf{q}_{1},\ldots ,\mathbf{q}_{K}\right] \in 
\mathbb{C}
^{J_{\mathrm{T}}\times K}$ is the transmit beamforming matrix, $\mathbf{q}%
_{k}\in \mathbb{C}^{J_{\mathrm{T}}\times 1}$ is the beamforming vector for
the signal of $U_k$, and $\mathbf{x\triangleq }\left[
x_{1},\ldots ,x_{K}\right] ^{T}\in 
\mathbb{C}
^{K\times 1}$ is the unit-power confidential signal vector of the set of users with a covariance matrix $\mathbf{R}_{x}=\mathbb{E}\left[ \mathbf{xx}^{H}\right] =%
\mathbf{I}_{K}$. Additionally, the AN$\ $signal $\mathbf{z}\in 
\mathbb{C}
^{J_{\mathrm{T}}\times 1}$, multiplexed along with the information signal,
is aimed to be projected onto the malicious targets' channel subspace, and
serves for a two-fold purpose:

\begin{itemize}
\item Degrading the malicious targets' decoding capabilities by decreasing
their received signal-to-interference-and-noise ratios (SINRs),

\item Increasing the sensing illumination power, e.g., improving the target
detection probability.
\end{itemize}

Such a signal is modeled as a zero-mean complex Gaussian process with
covariance matrix $\mathbf{R}_{z}=\mathbb{E}\left[ \mathbf{zz}^{H}\right] $.
Also, $n_{\Omega}$ is the reception zero-mean additive white Gaussian noise
(AWGN)\ at either a user or a malicious target of variance $\sigma
_{\Omega}^{2}$. As a result, based on the received signal expression in %
\eqref{rxsig1},\ one can formulate the received SINR\ at either $D_{k}$ and $%
T_{l}$ for decoding the signal $x_{k}$ as%
\begin{align}
\gamma _{\Omega }^{(k)} &=\frac{\mathbb{E}\left[ \left\vert \mathbf{h}%
_{S\Omega }\mathbf{q}_{k}\right\vert ^{2}\right] }{\mathbb{E}\left[%
\sum\limits_{k^{\prime }=1,k^{\prime }\neq k}^{K}\left\vert \mathbf{h}%
_{S\Omega }\mathbf{q}_{k^{\prime }}\right\vert ^{2}+ \left\vert \mathbf{h}%
_{S\Omega}\mathbf{z}\right\vert ^{2}\right] +\sigma _{\Omega }^{2}}  \notag
\\
&=\frac{\mathrm{Tr}\left[ \mathbf{G}_{S\Omega }\overline{\mathbf{Q}}_{k}%
\right] }{\sum\limits_{k^{\prime }=1,k^{\prime }\neq k}^{K}\mathrm{Tr}\left[ 
\mathbf{G}_{S\Omega }\overline{\mathbf{Q}}_{k^{\prime }}\right] +\mathrm{Tr}%
\left[ \mathbf{G}_{S\Omega }\mathbf{R}_{z}\right] +\sigma _{\Omega }^{2}},
\label{snruj}
\end{align}%
where $\mathbf{G}_{S\Omega}\triangleq \mathbf{h}_{S\Omega}^{H}\mathbf{h}%
_{S\Omega}$ and $\overline{\mathbf{Q}}_{k}\triangleq \mathbf{q}_{k}\mathbf{q}%
_{k}^{H}$.

\subsection{Key Performance Indicators}

\subsubsection{Eavesdropping Assumptions and Secrecy Evaluation Metric}

The SC metric assesses the theoretical secrecy limits of a wireless network
under eavesdropping attacks. SC is identified as the maximum communication
rate set by the legitimate parties to achieve (i)\ a reliable message
decoding at each legitimate receiver with an arbitrarily small decoding
error and (ii)\ a unit equivocation rate at the eavesdroppers. By
considering an infinite blocklength regime, one can formulate the
instantaneous SC as $C_{\mathrm{S}}=[C_{L}-C_{E}]^{+}$ where $%
[x]\triangleq \max(0,x)$, $C_{L}\triangleq \log _{2}\left( 1+\gamma
_{L}\right) $ and $C_{E}\triangleq \log _{2}\left( 1+\gamma _{E}\right) $
are defined as the legitimate and illegitimate channels' capacities,
respectively, expressed in terms of the respective links' instantaneous
SINRs. In the considered network, each of the $L$ malicious
targets/eavesdroppers attempts independently to illegally decode each user's
signal, i.e., non-colluding eavesdroppers, which results in the following\
expression for the analyzed system's SC: 
\begin{equation}
C_{\mathrm{S}}=\min_{\substack{ k=1,\ldots ,K  \\ L=1,\ldots ,L}}\left[
C_{D_{k}}^{(k)}-C_{T_{l}}^{(k)}\right] ^{+}.  \label{Csdl}
\end{equation}%
In \eqref{Csdl}, $C_{\mathrm{S}}$ defines the worst-case (minimal)\ SC out
of the various individual SCs linking the decoding of each confidential
signal and each eavesdropper, where 
\begin{equation}
C_{\Omega }^{(k)}=\log _{2}\left( 1+\gamma _{\Omega }^{(k)}\right) ,\Omega
\in \left\{ D_{k},T_{l}\right\}  \label{ccdluser}
\end{equation}%
represents the channel capacity of either the $k$th\ user channel or the $l$%
th eavesdropper one to decode $x_{k}$. Therefore, a secure system focuses on
maximizing the worst-case SC level or maintaining it above a fixed positive
threshold rate.

\subsubsection{Sensing Model Evaluation Metric}

In addition to securely serving the various DL\ users and preserving
information legitimacy in the considered network, another objective of the
DFRC\ BS is to sense the presence of the $L$ malicious targets by means of
monostatic radar sensing, relying on the arrays of $J_{\mathrm{T}}$ transmit
antennas and $J_{\mathrm{R}}$ receive ones.

One should emphasize that detecting the $L$ targets requires pointing
electromagnetic signal beams into the directions of the targets. Thus, this
assumes knowledge of their locations beforehand. Such location information
can be easily obtained by an initial sensing-aided localization phase, which
enables retrieving physical location parameters, such as the azimuth angle,
elevation angle, and the range of each target.

Accordingly, a robust target detection can be ensured by efficient\ transmit
beamforming design at $S$ and\ BD-RIS\ scattering at $R$ to maximize the
amount of electromagnetic power hitting each target \cite[Chapter 8]%
{emilbook}. To this end, the BD-RIS beampattern represents a useful metric
to evaluate the directivity of the signal beams reflected through the
various elements of $R$. Such a metric can be expressed as 
\begin{eqnarray}
V\left( \theta _{0},\phi _{0}\right) &=&\mathbb{E}\left[ \left\vert \mathbf{h%
}_{R}\left( \theta _{0},\phi _{0}\right) \mathbf{v}\right\vert ^{2}\right] 
\notag \\
&=&\mathbf{h}_{R}\left( \theta _{0},\phi _{0}\right) \mathbf{R}_{v}\mathbf{h}%
_{R}^{H}\left( \theta _{0},\phi _{0}\right) ,  \label{beampattern}
\end{eqnarray}%
where $\mathbf{h}_{R}\left( \theta _{0},\phi _{0}\right) \triangleq \mathbf{e%
}\left( \theta _{0},\phi _{0}\right) \mathbf{\Psi H}_{SR}$ and $\mathbf{R}%
_{v}\triangleq \mathbf{R}_{z}+\sum_{k=1}^{K}\overline{\mathbf{Q}}_{k}$ is
the transmit signal covariance matrix. In \eqref{beampattern}, $\theta _{0}$
and $\phi_{0}$ stands for the elevation and azimuth look directions from $R$%
's observation plane, where $V\left( \theta _{0},\phi _{0}\right) $
quantifies the amount of signal power beamsteered in the three-dimensional
direction defined by the pair of angles $\left( \theta _{0},\phi _{0}\right) 
$.

In addition to the BD-RIS beampattern, the per-target reflected power is
considered as another metric for evaluating sensing efficacy. Such a
quantity assesses the level of signal power reflected by each target \cite%
{illi2025fdris}, which can be formulated as 
\begin{eqnarray}
\mathcal{V}_{s}^{(l)} &=&\left\vert \overline{\mathbf{h}}_{ST_{l}}\mathbf{v}%
\right\vert ^{2}  \notag \\
&=&\mathrm{Tr}\left[ \overline{\mathbf{H}}_{ST_{l}}\mathbf{R}_{v}\right] ,
\label{senspw}
\end{eqnarray}%
where 
\begin{equation}
\overline{\mathbf{h}}_{ST_{l}}=\sqrt{\overline{\zeta }_{ST_{l}}}\mathbf{h}%
_{RT_{l}}\mathbf{\Psi H}_{SR}\in \mathbb{C}^{1\times J_{\mathrm{T}}}
\label{hstlbar}
\end{equation}%
represents the sensing channel vector between $S$ and $T_{l}$,%
\begin{equation}
\overline{\zeta }_{ST_{l}}=\frac{\mathcal{G}_{S,\mathrm{T}}\rho
_{R}^{4}\sigma _{\mathrm{RCS}}^{(l)}}{d_{SR}^{2}d_{RT_{l}}^{2}4\pi }
\label{fsplsens}
\end{equation}%
denotes the sensing FSPL term corresponding to the $l$th target, taking into
account its effective aperture represented by the radar cross section (RCS) $%
\sigma_{\mathrm{RCS}}^{(l)}$.\ Also, $\overline{\mathbf{H}}%
_{ST_{l}}\triangleq \overline{\mathbf{h}}_{ST_{l}}^{H}\overline{\mathbf{h}}%
_{ST_{l}}$.

\begin{remark}
Observe from \eqref{senspw}-\eqref{fsplsens} that the reflected power by
each target is mainly controlled by its RCS, the RE size (length of $\rho
_{R}$), the $S$-$R$ and $R$-$T_{l}$ distances, in addition to the BD-RIS\
scattering matrix $\mathbf{\Psi }$ and the ISAC\ signal beamforming $\mathbf{%
R}_{v}$. While $\sigma _{\mathrm{RCS}}^{(l)}$ and $d_{RT_{l}}$ are tied to
the physical properties (i.e., range, effective area, and look direction) of
the sensed target, $\rho _{R}$ and $\mathbf{\Psi }$ represent BD-RIS\ design
parameters which have a control over the beampattern profile of the\ BD-RIS.
Furthermore, the per-target reflection power is impacted by the beamforming
profile from the DFRC BS, controlled by the ISAC\ signal covariance matrix $%
\mathbf{R}_{v}$.
\end{remark}

\section{Optimization Framework and Algorithmic Solution Procedure}

In this section, the optimization problem at hand is presented and detailed.
In particular, the considered optimization problem aims at ensuring a maximal sensing performance of the analyzed system, subject to secrecy and total power constraints. Additionally, a detailed formulation of the proposed solution is given.

\subsection{Optimization Problem Description}

From a secure ISAC network perspective, the objective is three-fold, namely
ensuring (i)\ a reliable communication with the legitimate downlink users,
(ii)\ a minimal legitimate signal leakage to the malicious nodes, and (iii)
a robust sensing by maximizing the reflected power by each target. Thus,
this can be achieved by designing an optimal BD-RIS\ scattering matrix $%
\mathbf{\Psi }$, transmit beamforming vectors $\left\{ \mathbf{q}%
_{k}\right\}_{k=1}^{K}$, and AN\ covariance matrix $\mathbf{R}_{z}$. In the
studied system, sensing is prioritized over secrecy. Thus, a weighted linear
combination of the reflected signal power by the various targets is
considered as an objective function. Furthermore, the network's secrecy is
considered in the constraints, in addition to a total power constraint. The
BD-RIS's controller is linked to $S$ via a wired backhaul link, enabling
centralized optimization of the aforementioned control variables at $S$.
Also, a perfect CSI is assumed to be\ available at the DFRC BS, which is
aligned with the assumption of stationary nodes in the network and
sufficiently longer training pilot symbols for channel estimation \footnote{%
As the considered network's objective is performing target detection, it is
assumed that the locations of the various targets are known at $S$
beforehand. Locations can be estimated via an initial beam scanning-based
localization. Also, without loss of generality, a pure LoS\ link is
considered between $R$ and $\left\{ T_{l}\right\} _{l=1}^{L}$. Thus, one can
construct the CSI of the malicious targets, acting as passive devices, by
leveraging the estimated location of each target, producing an estimate of
the FSPL term in \eqref{fsplris}. Consequently, the channel response $%
\mathbf{h}_{ST_{l}}$ can be constructed utilizing the knowledge of $\mathbf{H%
}_{SR}$.}. As a result, the proposed optimization problem can be expressed
as 
\begin{subequations}
\label{PA}
\begin{align}
\mathcal{P}1& :\max_{\left\{ \overline{\mathbf{Q}}_{k}\right\} _{k=1}^{K}%
\mathbf{,R}_{z},\mathbf{\Psi }}\sum\limits_{l=1}^{L}\alpha _{l}\mathcal{V}%
_{s}^{(l)} \\
\text{s.t.}\ (\mathrm{C1})& :C_{\mathrm{S}}\geq C_{\mathrm{th}},  \label{C1a}
\\
(\mathrm{C2})& :\mathrm{Tr}\left[ \sum\limits_{k=1}^{K}\overline{\mathbf{Q}}%
_{k}+\mathbf{R}_{z}\right] \leq P_{\mathrm{S}}  \label{C2a} \\
(\mathrm{C3})& :\overline{\mathbf{Q}}_{k}\mathbf{\succeq 0}\text{ }\left(
k=1,\ldots ,K\right) \mathbf{,R}_{z}\mathbf{\succeq 0}  \label{C3a} \\
(\mathrm{C4})& :\mathrm{rank}\left( \overline{\mathbf{Q}}_{k}\right)
=1,k=1,\ldots ,K  \label{C4a} \\
(\mathrm{C5})& :\mathbf{\Psi }^{H}\mathbf{\Psi }=\mathbf{I}_{M}  \label{C5a}
\end{align}
\label{P1}
\end{subequations}
Observe from \eqref{P1} that the optimization problem can be controlled by
the rank-one positive semidefinite beamforming matrices $\{ \overline{%
\mathbf{Q}}_{k}\} _{k=1}^{K}$, which are linked to the beamforming
vectors $\left\{ \mathbf{q}_{k}\right\} _{k=1}^{K}$ of the various users. On
one hand, the objective function is expressed as a weighted sum of the
reflected powers by the $L$ targets, whereby $\left\{ \alpha _{l}\right\}
_{l=1}^{L}$ are arbitrarily-chosen weight coefficients tuned to prioritize
the increase of the sensing performance for a given set of targets out of
the $L$ ones. On the other hand, notice that the system's secrecy is defined
by $\mathrm{C1}$, where $C_{\mathrm{th}}$ is a threshold (minimal) SC level
set to ensure a secure transmission. Additionally, $\mathrm{C2}$ defines the
system's total power constraint, with $P_{\mathrm{S}}$ denoting $S$'s power
budget, $\mathrm{C3}$ forces a positive semidefinitness property on $\mathbf{%
R}_{z}$ and $\{ \overline{\mathbf{Q}}_{k}\} _{k=1}^{K} $ per
their definitions, and $\mathrm{C4}$ represents the rank-one property of $%
\{ \overline{\mathbf{Q}}_{k}\} _{k=1}^{K}$. Finally, $\mathrm{C5}$
represents the unitary nature of the BD-RIS scattering matrix. The problem
in \eqref{PA} presents several challenges, namely:

\begin{enumerate}
\item The per-target reflected power as well as the SC\ are bilinear in
either $\{ \overline{\mathbf{Q}}_{k}\} _{k=1}^{K}$ and $\mathbf{%
\Psi }$ or $\mathbf{R}_{z}\ $and $\mathbf{\Psi }$, as noted from \eqref{hsomega}, \eqref{snruj}-\eqref{ccdluser}, \eqref{senspw}, and \eqref{hstlbar}. This
results in the objective function and the constraint $\mathrm{C1}$ of $%
\mathcal{P}1$ to be non-convex jointly in the aforementioned variables.
Notably, one can note that $\{\overline{\mathbf{Q}}_{k}\}
_{k=1}^{K}$ are not coupled with $\mathbf{R}_{z}$.

\item For a given value of either $\{ \{ \overline{\mathbf{Q}}%
_{k}\} _{k=1}^{K},\mathbf{R}_{z}\} $ or $\mathbf{\Psi }$, the SC
constraint $\mathrm{C1}$ in \eqref{C1a} is non-convex due to the presence of
fractional terms in either of the two aforementioned control variable sets.

\item The rank-one constraints in \eqref{C4a} as well as the BD-RIS
scattering matrix unitary property in \eqref{C5a} are non-convex.
\end{enumerate}

\subsection{Alternative Convex Representation}

Initially, $\mathcal{P}1$ can be reformulated into the following form 
\begin{subequations}
\label{PB}
\begin{align}
\mathcal{P}2& :\max_{\left\{ \overline{\mathbf{Q}}_{k}\right\} _{k=1}^{K}%
\mathbf{,R}_{z},\mathbf{\Psi }}\sum\limits_{l=1}^{L}\alpha _{l}\mathcal{V}%
_{s}^{(l)}  \label{objb} \\
\text{s.t.}\ (\mathrm{C1})& :\gamma _{D_{k}}^{(k)}\geq \gamma _{D}^{(\min
)},\forall k,  \label{C1b} \\
(\mathrm{C2})& :\gamma _{T_{l}}^{(k)}<\gamma _{T}^{(\max )},\forall k,l
\label{C2b} \\
& \eqref{C2a}-\eqref{C5a}
\end{align}
\label{P2}
\end{subequations}

Note that in \eqref{PB}, the original SC\ constraint in \eqref{C1a} is
transformed into the SINR-based constraints in \eqref{C1b} and \eqref{C2b},
where $\gamma _{D}^{(\min )}$ is defined as the minimal legitimate SINR
while $\gamma _{T}^{(\max )}$ is the maximal allowed receive SINR at the
malicious targets. From the latter pair of equations, one can observe the
equivalence between $\mathcal{P}1$ and $\mathcal{P}2$, whereby we can define 
$C_{D}^{(\min )}\triangleq \log _{2}( 1+\gamma _{D}^{(\min )}) $
as the minimal legitimate channel capacity, while $C_{T}^{(\max )}\triangleq
\log _{2}( 1+\gamma _{T}^{(\max )}) $ denotes the maximal
eavesdropping channel capacity. Therefore, by virtue of the monotonicity of
the logarithm function, the inequalities of \eqref{C1b} and \eqref{C2b}
equivalently represent a lower bound for the legitimate channel capacity and
an upper bound for the wiretap channel, respectively, i.e., $%
C_{D_{k}}^{(k)}\geq C_{D}^{(\min )}$ and $C_{T_{l}}^{(k)}<C_{T}^{(\max )}$.
Thus, this produces the following 
\begin{equation}
\underset{\triangleq C_{\mathrm{S}}^{\left( k,l\right) }}{\underbrace{%
C_{D_{k}}^{(k)}-C_{T_{l}}^{(k)}}}\geq C_{\mathrm{th}},\forall k,l,
\label{secineq}
\end{equation}%
where $C_{\mathrm{th}}\triangleq C_{D}^{(\min )}-C_{T}^{(\max )}$. Note
that, according to the worst-case SC\ definition in (\ref{Csdl}), (\ref%
{secineq}) can be equivalently expressed as $C_{\mathrm{S}}\geq C_{\mathrm{th%
}}$ for $C_{\mathrm{S}}^{\left( k,l\right) }\geq 0$ values, which is exactly
the original secrecy constraint in \eqref{C1a}. Therefore, by leveraging
such an equivalent secrecy constraint representation in \eqref{C1b} and %
\eqref{C2b}, the minimal SC\ level can be tuned by modifying either the
minimal legitimate decoding SINR $\gamma _{D}^{(\min )}$ or the maximal
eavesdropping one $\gamma _{T}^{(\max )}$. Furthermore, observe that %
\eqref{C1b} and \eqref{C2b} can help in transforming the initial SC
constraint in \eqref{C1a} into a convex expression in terms of $\{
\{ \overline{\mathbf{Q}}_{k}\} _{k=1}^{K}\mathbf{,R}_{z}\} $
for a given $\mathbf{\Psi }$, or vice-versa. Thus, this addresses the second
challenge among the three challenges raised in the past subsection.

On the other hand, in order to circumvent the high coupling between $\{
\{ \overline{\mathbf{Q}}_{k}\} _{k=1}^{K}\mathbf{,R}_{z}\} $
and $\mathbf{\Psi }$ in \eqref{C1b} and \eqref{C2b}, an AO-based approach
can be utilized, in which $\mathcal{P}2$ in \eqref{PB} is split into two
subproblems and solved iteratively by optimizing in each subproblem one or
more non-coupled variables, while the rest of the variables are fixed. In the sequel, a detailed explanation of the considered AO-based optimization
is presented.

\subsection{Proposed Alternating Optimization-based Optimization Scheme}

\subsubsection{Optimal BD-RIS$\ $Scattering Matrix $\mathbf{\Psi }$ For
Given $\{ \overline{\mathbf{Q}}_{k}\} _{k=1}^{K}$\textbf{\ and }$%
\mathbf{R}_{z}$}

Observe that for given values of $\{ \overline{\mathbf{Q}}_{k}\}
_{k=1}^{K}$\textbf{\ }and\textbf{\ }$\mathbf{R}_{z}$, the problem in \eqref{P2} is
non-convex in terms of $\mathbf{\Psi }$. This is essentially due to the
quartic dependence of the per-target reflected power in \eqref{senspw}
and the SINRs in \eqref{snruj} on $\mathbf{\Psi }$, such as 
\begin{eqnarray}
\mathcal{V}_{s}^{(l)} &=&\left\vert \overline{\mathbf{h}}_{ST_{l}}\mathbf{v}%
\right\vert ^{2},  \notag \\
&=&\underset{\triangleq \overline{f}_{l}\left( \mathbf{\Psi },\left\{ 
\overline{\mathbf{Q}}_{k}\right\} _{k=1}^{K},\mathbf{R}_{z}\right) }{%
\underbrace{\mathrm{Tr}\left[ \overline{\mathbf{G}}_{RT_{l}}\mathbf{\Psi H}%
_{SR}\mathbf{R}_{v}\mathbf{H}_{SR}^{H}\mathbf{\Psi }^{H}\right] }}
\end{eqnarray}%
and%
\begin{equation}
\gamma _{\Omega }^{(k)}=\frac{\mathrm{Tr}\left[ \mathbf{G}_{R\Omega }\mathbf{%
\Psi H}_{SR}\overline{\mathbf{Q}}_{k}\mathbf{H}_{SR}^{H}\mathbf{\Psi }^{H}%
\right] }{\left[ 
\begin{array}{c}
\sum\limits_{k^{\prime }=1,k^{\prime }\neq k}^{K}\mathrm{Tr}\left[ \mathbf{G}%
_{R\Omega }\mathbf{\Psi H}_{SR}\overline{\mathbf{Q}}_{k^{\prime }}\mathbf{H}%
_{SR}^{H}\mathbf{\Psi }^{H}\right]  \\ 
+\mathrm{Tr}\left[ \mathbf{G}_{R\Omega }\mathbf{\Psi H}_{SR}\mathbf{R}_{z}%
\mathbf{H}_{SR}^{H}\mathbf{\Psi }^{H}\right] +\sigma _{\Omega }^{2}%
\end{array}%
\right] },
\end{equation}%
for $\Omega \in \left\{ D_{k},T_{l}\right\} $, where $\overline{\mathbf{G}}%
_{RT_{l}}\triangleq \overline{\zeta }_{ST_{l}}\mathbf{h}_{RT_{l}}^{H}\mathbf{%
h}_{RT_{l}}$ and $\mathbf{G}_{R\Omega }\triangleq \zeta _{S\Omega }\mathbf{h}%
_{R\Omega }^{H}\mathbf{h}_{R\Omega }$. In addition, another challenge in $%
\mathcal{P}2$ lies in the non-convex unitary property constraint in %
\eqref{C5a}. To tackle this cumbersome problem, we resort to a Riemannian
manifold optimization-based procedure for optimizing the BD-RIS scattering
matrix. Therefore, the AL corresponding to $\mathcal{P%
}2$ can be formulated as \cite{boumal2}  
\begin{align}
\mathcal{J}\left( \mathbf{\Psi },\bm{\upbeta},\mathbf{D}\right) & =-f\left( 
\mathbf{\Psi },\mathbf{D}\right)  \notag \\ & +\frac{\varrho }{2}\sum\limits_{i=1}^{K%
\left( L+1\right) }\left\{ \max \left[ 0,\frac{\beta _{i}}{\varrho }%
+g_{i}\left( \mathbf{\Psi },\mathbf{D}\right) \right] \right\} ^{2},
\label{lagrangian}
\end{align}%
where $\mathbf{D}\triangleq \{\{\overline{\mathbf{Q}}_{k}\}_{k=1}^{K},%
\mathbf{R}_{z}\}$, $f(\mathbf{\Psi },\mathbf{D})\triangleq \sum_{l=1}^{L}%
\alpha_l \overline{f}_{l}\left( \mathbf{\Psi },\mathbf{D}\right) $ represents the objective function (i.e., sum reflected power). Since the objective of the original problem in \eqref{P1} is to maximize this function, a minus sign is introduced so that it can be handled as a minimization problem within the AL framework. Also, $\varrho >0$ is a penalty parameter, and $g_{i}(\mathbf{\Psi },\mathbf{D})$ is defined in %
\eqref{giconst} at the top of the page, 
\begin{figure*}[t]
{\normalsize 
\setcounter{mytempeqncnt}{\value{equation}} 
}
\par
\begin{equation}
g_{i}\left( \mathbf{\Psi },\mathbf{D}\right) =\left\{ 
\begin{array}{l}
\gamma _{{D}}^{(\min )}\left[ \sum\limits_{k^{\prime }=1,k^{\prime }\neq
k}^{K}\mathrm{Tr}\left[ {\mathbf{G}}_{RD_{i}}\mathbf{\Psi H}_{SR}\overline{%
\mathbf{Q}}_{k^{\prime }}\mathbf{H}_{SR}^{H}\mathbf{\Psi }^{H}\right] +%
\mathrm{Tr}\left[ {\mathbf{G}}_{RD_{i}}\mathbf{\Psi H}_{SR}\mathbf{R}_{z}%
\mathbf{H}_{SR}^{H}\mathbf{\Psi }^{H}\right] +\sigma _{D_{i}}^{2}\right] \\ 
-\mathrm{Tr}\left[ {\mathbf{G}}_{RD_{i}}\mathbf{\Psi H}_{SR}\overline{%
\mathbf{Q}}_{k}\mathbf{H}_{SR}^{H}\mathbf{\Psi }^{H}\right] ,i\leq K \\ 
\mathrm{Tr}\left[ {\mathbf{G}}_{RT_{\left\lfloor \left( i-1\right)
/K\right\rfloor }}\mathbf{\Psi H}_{SR}\overline{\mathbf{Q}}_{\left(
i-K-1\right) \mathrm{mod}K}\mathbf{H}_{SR}^{H}\mathbf{\Psi }^{H}\right] \\ 
-\gamma _{{T}}^{(\max )}\left[ 
\begin{array}{c}
\sum\limits_{k^{\prime }=1,k^{\prime }\neq k}^{K}\mathrm{Tr}\left[ {\mathbf{G%
}}_{RT_{\left\lfloor \left( i-1\right) /K\right\rfloor }}\mathbf{\Psi H}_{SR}%
\overline{\mathbf{Q}}_{k^{\prime }}\mathbf{H}_{SR}^{H}\mathbf{\Psi }^{H}%
\right] \\ 
+\mathrm{Tr}\left[ {\mathbf{G}}_{RT_{\left\lfloor \left( i-1\right)
/K\right\rfloor }}\mathbf{\Psi H}_{SR}\mathbf{R}_{z}\mathbf{H}_{SR}^{H}%
\mathbf{\Psi }^{H}\right] +\sigma _{T_{\left\lfloor \left( i-1\right)
/K\right\rfloor }}^{2}%
\end{array}%
\right] ,K<i\leq K\left( L+1\right)%
\end{array}%
\right.  \label{giconst}
\end{equation}%
\par
{\normalsize 
\hrulefill 
\vspace*{1pt} }
\end{figure*}
Also, $\beta _{i}\geq 0$ denote the Lagrangian dual (multipliers)\
accounting for the actual weight of the $i$th constraint, with $\bm{\upbeta}\triangleq%
\left[ \beta _{1},\ldots ,\beta _{K\left( L+1\right) }\right] $. On the one
hand, since the AL in \eqref{lagrangian} models the optimization with
respect to $\mathbf{\Psi }$ from the primal problem \eqref{PB}, the
constraints in \eqref{C2a}-\eqref{C4a} are not incorporated in $\mathcal{J}%
\left( \mathbf{\Psi },\bm{\upbeta},\mathbf{D}\right) $ due to their
independence of the BD-RIS\ scattering matrix $\mathbf{\Psi }$. On the other
hand, note that $\mathbf{\Psi }$ lies in $\mathrm{St}(M,M)$. Thus, while the
unitary property constraint on $\mathbf{\Psi }$ in \eqref{C5a} is dropped
from $\mathcal{J}\left( \mathbf{\Psi },\bm{\upbeta},\mathbf{D}\right) $, it
can be taken into consideration by minimizing the AL over a Stiefel
manifold, guaranteeing the unitarity of the achieved minima of $\mathcal{J}(%
\mathbf{\Psi },\bm{\upbeta},\mathbf{D})$.

First, one can note that due to the unitary property of $\mathbf{\Psi }$, it
can be formulated as a product of a given BD-RIS\ scattering matrix (e.g.,
suboptimal guess), denoted by $\mathbf{\Psi }_{0}$, and a unitary matrix $%
\mathbf{X}$, i.e., $\mathbf{\Psi =\Psi }_{0}\mathbf{X}$. Such a
reformulation renders the optimization strategy's focus on identifying the
optimal unitary rotation matrix $\mathbf{X}$ starting from an initial guess
of $\mathbf{\Psi }$, i.e., $\mathbf{\Psi }_{0}$. For instance, the latter
can be set to be aligned with the channel matrix of the $S$-$R$ link.
Therefore, as the rotation of matrices in $\mathrm{St}(M,M)$ by unitary matrices maintains them under the same manifold, it can facilitate a sequential
optimization of $\mathbf{\Psi }$ as:\ $\mathbf{\Psi }_{m}=\mathbf{\Psi }%
_{m-1}\mathbf{X}$, where the superscript $m$ refers to the current
optimization iteration and $\mathbf{\Psi }_{m}$ is the solution obtained at
iteration $m$. Thus, the equivalent optimization problem on $\mathbf{\Psi }$
can be reformulated in terms of $\mathbf{X}$, $\mathbf{\Psi}_{m-1}$, and $m$
as 
\begin{align}
\mathcal{P}3^{(m)}:& \min_{\mathbf{X}}\mathcal{J}^{(m)}\left( \mathbf{X,%
\mathbf{\Psi }}_{m-1}, \bm{\upbeta}^{(m)}, \mathbf{D}_{m-1}\right)  \notag \\
\text{s.t. }& \mathbf{X}\mathbf{\in }\mathrm{St}\left( M,M\right)
\label{P3m}
\end{align}%
with $\mathcal{J}^{(m)}\left( \mathbf{X,\mathbf{\Psi }}_{m-1},%
\bm{\upbeta}^{(m)} ,\mathbf{D}_{m-1}\right) $ defining the reformulated AL in \eqref{lagrangian} by
substituting $\mathbf{\Psi }$ with\textbf{\ }$\mathbf{\mathbf{\Psi }}_{m-1}%
\mathbf{X}$. Note that a index $(m-1)$ is added to $\mathbf{D}$ to indicate that the considered value of $\mathbf{D}$ at the current ($m$th) iteration, obtained from the past iteration. Furthermore, the superscript $m$ is added to the Lagrangian dual
vector $\bm{\upbeta}$ to indicate the current iteration's considered
constraint multipliers. Thus, for a given $\bm{\upbeta }^{(m)}$, an RCG
algorithm is employed to solve the unconstrained problem in \eqref{P3m}
through the following steps \cite{Boumaljournal}:

\begin{enumerate}
\item \textbf{Euclidean Gradient Formulation:} First, the Euclidean gradient
of the Lagrangian is computed with respect to $\mathbf{X}$ as 
\begin{align}
& \nabla _{\mathrm{E,}\mathbf{X}}\mathcal{J}^{(m)}\left( \mathbf{X},\mathbf{
\Psi }_{m-1},\bm{\upbeta}^{(m)},\mathbf{D}_{m-1}\right)   \notag \\
& =-\nabla _{\mathrm{E}}f^{(m)}\left( \mathbf{X},\mathbf{
\Psi }_{m-1},\mathbf{D}_{m-1}\right)   \notag \\
& +\frac{\varrho_m }{2}\sum\limits_{i=1}^{K\left( L+1\right) }\max \left[ 0,%
\frac{\beta _{i}}{\varrho_m }+g_{i}^{(m)}\left( \mathbf{X},\mathbf{
\Psi }_{m-1} ,\mathbf{D}_{m-1}\right) \right]   \notag \\
& \times \nabla _{\mathrm{E,}\mathbf{X}}g_{i}^{(m)}\left( \mathbf{X},\mathbf{
\Psi }_{m-1}, \mathbf{D}_{m-1}\right)   \label{gradlagrangian}
\end{align}%
where $f^{(m)}(.)$ and $g_{i}^{(m)}(.)$ are the equivalent form of $f(.)$
and $g_{i}(.)$ in \eqref{lagrangian} by substituting $\mathbf{\Psi }$ with%
\textbf{\ }$\mathbf{\mathbf{\Psi }}_{m-1}\mathbf{X}$, whereas the index $m$ was appended to the penalty factor $\varrho$ in order to account for the current weight of the set of constraints at the $m$th iteration. Also, $%
\mathbf{D}_m\triangleq \{\{\overline{\mathbf{Q}}_{k}^{(m)}\}_{k=1}^{K},\mathbf{R}_{z}^{(m)}\}$ denotes the obtained solutions at
the $m$th iteration for $\{\{ \overline{\mathbf{Q}}_{k}\}
_{k=1}^{K},\mathbf{R}_{z}\}$, whereas 
\begin{align}
\nabla _{\mathrm{E,}\mathbf{X}}f^{(m)}\left( \mathbf{X},\mathbf{
\Psi }_{m-1}, \mathbf{D}_{m-1}\right) & =2\frac{\partial f^{(m)}\left( \mathbf{X},\mathbf{
\Psi }_{m-1}, \mathbf{D}_{m-1}\right) }{\partial \mathbf{X}^{\ast }}  \notag \\
& =2\sum_{l=1}^{L}\mathbf{\Psi }^{H}\overline{\mathbf{G}}_{RT_{l}}\mathbf{%
\Psi }  \notag \\
& \times \mathbf{XH}_{SR}\mathbf{R}_{v}\mathbf{H}_{SR}^{H},  \label{gradobj}
\end{align}%
and $\nabla _{\mathrm{E,}\mathbf{X}}g_{i}^{(m)}$, shown in \eqref{gradconst}
at the top of the next page, is computed similarly to $\nabla _{\mathrm{E,}%
\mathbf{X}}f^{(m)}$. 
\begin{figure*}[t]
{\normalsize 
\setcounter{mytempeqncnt}{\value{equation}} 
}
\par
\begin{equation}
\nabla _{\mathrm{E,}\mathbf{X}}g_{i}^{(m)}\left( \mathbf{X},\mathbf{\Psi }%
_{m-1},\mathbf{D}_{m-1}\right) =\left\{ 
\begin{array}{l}
2\left\{ \gamma _{D}^{(\min )}\left[ 
\begin{array}{c}
\sum\limits_{k^{\prime }=1,k^{\prime }\neq k}^{K}\mathbf{\Psi} _{m-1}^{H}{\mathbf{G}}%
_{RD_{i}}\mathbf{\Psi}_{m-1}\mathbf{X}\mathbf{H}_{SR}\overline{\mathbf{Q}}_{k^{\prime }}^{(m-1)}%
\mathbf{H}_{SR}^{H} \\ 
+\mathbf{\Psi} _{m-1}^{H}{\mathbf{G}}_{RD_{i}}\mathbf{\Psi} _{m-1}\mathbf{X}\mathbf{H}_{SR}\mathbf{R}_{z}^{(m-1)}%
\mathbf{H}_{SR}^{H}%
\end{array}%
\right] \right.  \\ 
-\left. \mathbf{\Psi }_{m-1}^{H}{\mathbf{G}}_{RD_{i}}\mathbf{\Psi }_{m-1}\mathbf{X}
\mathbf{H}_{SR}\overline{\mathbf{Q}}_{k}^{(m-1)}\mathbf{H}_{SR}^{H}\right\} ,i\leq K
\\ 
2\left\{ \mathbf{\Psi }_{m-1}^{H}{\mathbf{G}}_{RT_{\left\lfloor \left(
i-1\right) /K\right\rfloor }}\mathbf{\Psi }_{m-1}\mathbf{X}\mathbf{H}_{SR}\overline{%
\mathbf{Q}}_{\left( i-K-1\right) \mathrm{mod}K}^{(m-1)}\mathbf{H}_{SR}^{H}\right. 
\\ 
-\left. \gamma _{{T}}^{(\max )}\left[ 
\begin{array}{c}
\sum\limits_{k^{\prime }=1,k^{\prime }\neq k}^{K}\mathbf{\Psi }_{m-1}^{H}{%
\mathbf{G}}_{RT_{\left\lfloor \left( i-1\right) /K\right\rfloor }}\mathbf{%
\Psi }_{m-1} \mathbf{X} \mathbf{H}_{SR}\overline{\mathbf{Q}}_{k^{\prime }}^{(m-1)}\mathbf{H}%
_{SR}^{H} \\ 
+\mathbf{\Psi }_{m-1}^{H}{\mathbf{G}}_{RT_{\left\lfloor \left( i-1\right)
/K\right\rfloor }}\mathbf{\Psi }_{m-1}\mathbf{X}\mathbf{H}_{SR}\mathbf{R}_{z}^{(m-1)}\mathbf{H}%
_{SR}^{H}%
\end{array}%
\right] \right\} ,K<i\leq K\left( L+1\right) 
\end{array}%
\right.   \label{gradconst}
\end{equation}%
\par
{\normalsize 
\hrulefill 
\vspace*{1pt} }
\end{figure*}

\item \textbf{Transformation to a Riemannian Gradient:} As the Euclidean
gradient in \eqref{gradlagrangian} lies generally in the tangent space of a
point $\mathbf{X}$ in $%
\mathbb{C}
^{M\times M}$, a projection is required onto the tangent space of the
manifold, as shown in \eqref{gradriemlag} at the top of the page \cite%
{stiefelman}. Such a step is crucial in order to ensure a descent on $%
\mathbf{X}$ along the tangent space to the manifold. 
\begin{figure*}[t]
{\normalsize 
\setcounter{mytempeqncnt}{\value{equation}} 
}
\par
\begin{equation}
\nabla _{\mathrm{R,}\mathbf{X}}\mathcal{J}^{(m)}\left( \mathbf{X},\mathbf{\Psi }_{m-1} ,\bm{\upbeta}^{(m)} ,\mathbf{D}_{m-1}\right) =\nabla _{\mathrm{%
E,}\mathbf{X}}\mathcal{J}^{(m)}\left( \mathbf{X},\mathbf{\Psi }_{m-1} ,\bm{\upbeta}^{(m)} ,\mathbf{D}_{m-1}\right) -\mathbf{\mathbf{\mathbf{\mathbf{X}}}} %
\left[ \nabla _{\mathrm{E,}\mathbf{X}}\mathcal{J}^{(m)}\left( \mathbf{X},\mathbf{\Psi }_{m-1} ,\bm{\upbeta}^{(m)} ,\mathbf{D}_{m-1}\right) \right] ^{H}\mathbf{\mathbf{%
\mathbf{\mathbf{X}}}} .  \label{gradriemlag}
\end{equation}%
\par
{\normalsize 
\hrulefill 
\vspace*{1pt} }
\end{figure*}

\item \textbf{Iterative Gradient Descent:} Herein, a descent is performed on $\mathbf{X}$ from a given point $(\mathbf{X}_{m-1},\mathbf{D}_{(m-1)})$ over multiple
steps to reach a minimum of $\mathcal{J}^{(m)}(.)$, i.e., $\mathbf{X}_{m}$.
The gradient descent is performed iteratively according to the following
steps:

\begin{enumerate}
\item When using the RCG, the gradient at the current iteration needs to be
combined with previous step's search direction. Thus, a transportation
operation of the gradient at the current step needs to be performed before
combining it with the current step's gradient, i.e., 
\begin{align}
\mathbf{Y}_{\mathbf{X}}^{(m,n)}& =-\nabla _{\mathrm{R,}\mathbf{X}}\mathcal{J}%
^{(m)}\left( \mathbf{X}_{m}^{(n)},\mathbf{\mathbf{\Psi }}_{m-1},\bm{\upbeta}%
^{(m)},\mathbf{D}_{m-1}\right)   \notag \\
& +\varepsilon _{n}\mathrm{trans}_{\mathbf{X}_{m}^{(n)}\mathcal{\leftarrow }%
\mathbf{X}_{m}^{(n-1)}}\left( \mathbf{Y}_{\mathbf{X}}^{(m,n-1)}\right) 
\label{direction}
\end{align}%
where $\mathbf{Y}_{\mathbf{X}}^{(m,n)}$ is the search direction at the $n$th
gradient descent step and the $m$th AO iteration. In \eqref{direction}, the
index $n$ refers to the current ($n$th) gradient descent step. Without loss
of generality, a typical transportation corresponds to a projection of $%
\mathbf{Y}_{\mathbf{X}}^{(m,n-1)}$ into the tangent space at $\mathbf{X}%
_{m}^{(n)}$, denoted by $\mathcal{T}_{\mathbf{X}_{m}^{(n)}}\mathcal{M}$.
Note that for each AO iteration, one can set $\mathbf{X}_{m}^{(0)}=%
\mathbf{X}_{m-1}^{(N_{\mathrm{it}})}$, where $N_{\mathrm{it}}$ is the total
number of gradient descent steps upon reaching convergence in the inner RCG
loop, whereas 
\begin{equation}
\mathbf{Y}_{\mathbf{X}}^{(m,0)}=-\nabla _{\mathrm{R,}\mathbf{X}}\mathcal{J}%
^{(m)}\left( \mathbf{X}_{m}^{(0)},\mathbf{\mathbf{\Psi }}_{m-1},\bm{\upbeta}%
^{(m)},\mathbf{D}^{(m)}\right) .
\end{equation}%
Furthermore, $\varepsilon _{n}$ is an inertia parameter at the $n$th step,
chosen according to a given criterion. For instance, the Polak-Ribiere
formula can be utilized for inertia parameters as given by \eqref{epsilonn}
shown at the top of the page \cite{lit7}. 
\begin{figure*}[t]
{\normalsize 
\setcounter{mytempeqncnt}{\value{equation}} 
} 
\begin{equation}
\varepsilon _{n}=\frac{\mathrm{Tr}\left[ 
\begin{array}{c}
\left( 
\begin{array}{c}
\nabla _{\mathrm{R,}\mathbf{X}}\mathcal{J}^{(m)}\left( \mathbf{X}^{(n)}_{m}%
\mathbf{,\mathbf{\mathbf{\mathbf{\Psi }}}}_{m-1} ,\bm{\upbeta}^{(m)},\mathbf{%
D}_{m-1}\right) \\ 
-\nabla _{\mathrm{R,}\mathbf{X}}\mathcal{J}^{(m)}\left(\mathbf{X}^{(n-1)}_{m}%
\mathbf{,\mathbf{\mathbf{\mathbf{\Psi }}}}_{m-1} ,\bm{\upbeta}^{(m)},\mathbf{%
D}_{m-1}\right)%
\end{array}%
\right) \\ 
\times \left[ \nabla _{\mathrm{R,}\mathbf{X}}\mathcal{J}^{(m)}\left( \mathbf{%
X}^{(n)}_{m}\mathbf{,\mathbf{\mathbf{\mathbf{\Psi }}}}_{m-1} ,\bm{\upbeta}%
^{(m)},\mathbf{D}_{m-1} \right) \right] ^{H}%
\end{array}%
\right] }{\mathrm{Tr}\left[ \left( 
\begin{array}{c}
\nabla _{\mathrm{R,}\mathbf{X}}\mathcal{J}^{(m)}\left( \mathbf{X}^{(n-1)}_{m}%
\mathbf{,\mathbf{\mathbf{\mathbf{\Psi }}}}_{m-1} ,\bm{\upbeta}^{(m)},\mathbf{%
D}_{m-1} \right) \\ 
\times \left[ \nabla _{\mathrm{R,}\mathbf{X}}\mathcal{J}^{(m)}\left(\mathbf{X%
}^{(n-1)}_{m}\mathbf{,\mathbf{\mathbf{\mathbf{\Psi }}}}_{m-1} ,\bm{\upbeta}%
^{(m)},\mathbf{D}_{m-1} \right) \right] ^{H}%
\end{array}%
\right) \right] }  \label{epsilonn}
\end{equation}%
\par
{\normalsize 
\hrulefill 
\vspace*{1pt} }
\end{figure*}

\item The next gradient-based search point is generated, in an analogous way
to the conventional gradient descent-based methods, as follows 
\begin{equation}
\mathbf{X}_{m}^{(n+1)}=\mathrm{retr}\left( \mathbf{X}_{m}^{(n)}+\vartheta
_{n}\mathbf{Y}_{\mathbf{X}}^{(m,n)}\right) 
\end{equation}%
where $\vartheta _{n}$ is the gradient descent step size, which can be optimized
per the Armijo rule at each iteration. The retraction operation can be
defined by a thin singular-value decomposition, which enables moving along
the gradient-guided direction while remaining on the manifold, i.e.,\textrm{%
\ }$\mathrm{retr}\left( \mathbf{A}\right) =\mathbf{KL}^{H}$, with $\mathbf{KL%
}$ are defined such as: $\mathbf{A=K\Sigma L}^{H}$ and $\mathbf{\Sigma }$ is
the diagonal matrix of $\mathbf{A}$'s singular values.
\end{enumerate}

\item Once the $N_{\mathrm{it}}$ descent steps are performed, the BD-RIS
scattering matrix is updated by rotating it using $\mathbf{X}_{m}^{(N_{%
\mathrm{it}})}$ as: $\mathbf{\Psi }_{m}=\mathbf{\Psi }_{m-1}\mathbf{X}%
_{m}^{(N_{\mathrm{it}})}$.\newline
\end{enumerate}

\subsubsection{Update of the Lagrangian Multipliers and Penalty Factor}

The Lagrangian duals can be updated per the fulfillment of the problem's constraints, such as \cite{boumal2} 
\begin{equation}
\beta _{i}^{(m+1)}=\max \left( 0,\beta _{i}^{(m)}+\varrho_m g_{i}^{(m)}\left( \mathbf{X }_{m}^{(N_{\mathrm{it}})},
\mathbf{\Psi }_{m-1},\mathbf{D}_{m}\right) \right)    \label{betaa}
\end{equation}%
where the superscript $m+1$ on $\beta _{i}^{(m+1)}$ denotes the updated
value of the multiplier for the $(m+1)$th iteration's subproblem. Thus, the updated set of multipliers will be
taken into account in the AL of the upcoming AO iteration.

As far as the penalty coefficient $(\varrho_m)$ is concerned, its update is performed based on the degree of violation of the set of constraints. In particular, $\varrho$'s update rule is set as
\begin{equation}
    \varrho_{m+1}=\min\{ \delta \varrho_m ,  \varrho_{\max}\} \label{varrhoeq}
\end{equation}
for $\delta>1$, whenever the following condition is met at the $m$th iteration:
\begin{equation}
  \max_{i=1,\ldots,K(L+1)}  \left|\bar{g}_i^{(m)} \right| > \epsilon   \max_{i=1,\ldots,K(L+1)} \left|\bar{g}_i^{(m-1)} \right|, m>1,
\end{equation} with 
\begin{equation}
    \bar{g}_i^{(m)} \triangleq \max \left\{{g}_i^{(m)}\left( \mathbf{X }_{m}^{(N_{\mathrm{it}})},
\mathbf{\Psi }_{m-1},\mathbf{D}_{m}\right), -\beta^{(m)}_i/\varrho_m \right \}
\end{equation}
for $0<\epsilon<1$. Observe that the update of $\varrho_m$ in \eqref{varrhoeq} can either up-scale the penalty factor by a coefficient $\delta$ or clip it to a ceiling penalty coefficient value $(\varrho_{\max})$ in order to avoid an ill-conditioned AL \cite{illcond}.
\subsubsection{Optimal $\{ \overline{\mathbf{Q}}_{k}\} _{p=1}^{P}$%
\textbf{\ and }$\mathbf{R}_{z}$ For Given $\mathbf{\Psi }$}

For a given iteration of the iterative problem, and once the optimal BD-RIS
scattering matrix ($\mathbf{\Psi }_{m}$) is given, the subproblem $%
\mathcal{P}2$ becomes $\mathcal{P}4^{(m)}$ shown in \eqref{PB2} at the top
of the page after involving the expressions of $\gamma _{D_{k}}^{(k)}$ and $%
\gamma _{T_{l}}^{(k)}$ from \eqref{snruj} into \eqref{PB}. 
\begin{figure*}[t]
{\normalsize 
\setcounter{mytempeqncnt}{\value{equation}} 
}
\par
\begin{subequations}
\label{PB2}
\begin{align}
\mathcal{P}4^{(m)}& :\max_{\left\{ \overline{\mathbf{Q}}_{k}\right\}
_{k=1}^{K}\mathbf{,R}_{z}}\sum\limits_{l=1}^{L}\alpha _{l}\mathrm{Tr}\left[ 
\overline{\mathbf{H}}_{ST_{l}}\left( \sum\limits_{k=1}^{K}\overline{\mathbf{Q%
}}_{k}+\mathbf{R}_{z}\right) \right]  \label{objecB2} \\
\text{s.t.}\ (\mathrm{C1})& : \mathrm{Tr}\left[ \mathbf{G}_{SD_k }\overline{%
\mathbf{Q}}_{k}\right] - \gamma _{D}^{(\min )} \left( \sum\limits_{k^{\prime
}=1,k^{\prime }\neq k}^{K}\mathrm{Tr}\left[ \mathbf{G}_{SD_k }\overline{%
\mathbf{Q}}_{k^{\prime }}\right] +\mathrm{Tr}\left[ \mathbf{G}_{SD_k}\mathbf{%
R}_{z}\right] +\sigma _{D_k }^{2} \right) \geq 0,\forall k,  \label{C1b2} \\
(\mathrm{C2})& :\mathrm{Tr}\left[ \mathbf{G}_{ST_l }\overline{\mathbf{Q}}_{k}%
\right] - \gamma _{T}^{(\max )} \left( \sum\limits_{k^{\prime }=1,k^{\prime
}\neq k}^{K}\mathrm{Tr}\left[ \mathbf{G}_{ST_l }\overline{\mathbf{Q}}%
_{k^{\prime }}\right] +\mathrm{Tr}\left[ \mathbf{G}_{ST_l}\mathbf{R}_{z}%
\right] +\sigma _{T_l }^{2} \right) \leq 0,\forall k,l  \label{C2b2} \\
&\eqref{C2a}-\eqref{C4a}
\end{align}
\end{subequations}
\par
{\normalsize 
\hrulefill 
\vspace*{1pt} }
\end{figure*}
Note that the constraint \eqref{C5a}, related to the BD-RIS scattering matrix's unitarity, was dropped from $\mathcal{P}4^{(m)}$ due to its independence of the set of beamforming and AN covariance matrices. Thus, one
can observe that the problem in \eqref{PB2} contains convex objective and
constraint functions, i.e., \eqref{objecB2}-\eqref{C2b2}, \eqref{C2a}, as
they are expressed in terms of positive semidefinite matrices. On the other
hand, the rank-one constraint in \eqref{C4a} is the only non-convex one in %
\eqref{PB2}. To remedy this issue, an SDR is adopted, whereby the rank-one constraint is relaxed. As a result, $\mathcal{P%
}4^{(m)}$ becomes a convex semidefinite program (SDP) which can be
straightforwardly solved by any standard convex optimization solvers, e.g.,
CVX, MOSEK. Accordingly, due to the adoption of the SDR, the obtained
solution for $\{ \overline{\mathbf{Q}}_{k}\} _{k=1}^{K}$ may be
of a rank higher than one, which does not fulfill \eqref{C4a}. Thus, a practical rank-one solution can be obtained by virtue of the eigenvalue
decomposition (EVD) of $\overline{\mathbf{Q}}_{k}$ to represent it in terms
of its main eigenvector, i.e., $\overline{\mathbf{Q}}_{k}=\mathbf{B}\mathbf{%
\Lambda }\mathbf{B}^{H}$ \cite{illi2025fdris}. Therefore, the rank-one
solution for $\overline{\mathbf{Q}}_{k}$ at the $m$th iteration of the AO
framework is formulated as $\overline{\mathbf{Q}}_{k}^{(\mathrm{opt},m)}=%
\left[ \mathbf{\Lambda }\right] _{k^{\ast },k^{\ast }}\left[ \mathbf{B}%
\right] _{:,k^{\ast }}$, where $\mathbf{B}$ is a unitary matrix containing
the set of $J_{\mathrm{T}}$ eigenvectors, $\mathbf{\Lambda }$ is a diagonal
matrix containing $\overline{\mathbf{Q}}_{k}$'s eigenvalues, and 
\begin{equation}
k^{\ast }=\arg \max_{k=1,\ldots ,K}\left[ \mathbf{\Lambda }\right] _{k,k}
\end{equation}%
corresponds to the index of the largest eigenvalue of $\overline{\mathbf{Q}}%
_{k}$.
Algorithm \ref{algg} summarizes the above-detailed AO-based optimization framework for solving \eqref{P1}.

\begin{theorem}
    Let the set $\mathcal{X}=\left\{ \left.\mathbf{\Psi}, \mathbf{D} \right| \eqref{C1a}-\eqref{C5a} \text{ hold}\right\}$ represents the set of feasible points for $\mathcal{P}1$ in \eqref{P1}. The sequence of values $\mathbf{T}_m \triangleq \left\{  \mathbf{\Psi}_m, \mathbf{D}_m  \right\}$ produced by Algorithm \ref{algg} satisfies $f(\mathbf{T}_{m+1})\geq f(\mathbf{T}_m)$ for a sufficiently high $m$, and the objective function $f(\mathbf{T}_{m})$ converges asymptotically to a finite value $f^{\star}$.
    \begin{IEEEproof}
        First, observe that the reformulated problem in \eqref{P2} yields an equivalent representation of the original problem in \eqref{P1} by providing an alternative representation of the system's SC in terms of its legitimate and illegitimate SINRs, as detailed after \eqref{P2}. Therefore, the maximizer of the objective of the former (original) problem maximizes also the latter ($\mathcal{P}2$). 
        
        Furthermore, as far as the first AL-based suproblem is concerned, it should be noted that the minimizer $\mathbf{\Psi}^{\star}$ belongs to $\mathrm{St}(M,M)$, which is compact and closed \cite{Boumal_2023}. Thus, for a given feasible point $\mathbf{D}_{m-1}$ fed to the first AL-based subproblem, the RCG performs $N_{\mathrm{it}}$ descent steps to reach a minimizer for the AL in \eqref{lagrangian}. Herein, $\exists m_0$ such that $\varrho_{m}=\varrho_{m_0}$ for $m\geq m_0$, i.e., penalty factor stability. Thus, this takes place when 
            \begin{equation}
  \max_{i=1,\ldots,K(L+1)}  \left|\bar{g}_i^{(m)} \right| < \epsilon   \max_{i=1,\ldots,K(L+1)} \left|\bar{g}_i^{(m-1)} \right|. 
\end{equation} This means that $\max_{i=1,\ldots,K(L+1)}  \left|\bar{g}_i^{(m)} \right| \rightarrow 0$ and, consequently, $\max\{0,{g}_i^{(m)}(\mathbf{\Psi}_m,\mathbf{D}_{m-1}) \}\rightarrow0$ $(\forall i)$, i.e., constraints tend to be more fulfilled over iterations for a sufficiently large $\varrho_{m_0}$. This ensures the convergence to a minimizer $\mathbf{\Psi}^{\star}$ that decreases the objective function of the AL in \eqref{lagrangian}, i.e., $-f(\mathbf{\Psi},\mathbf{D}_m) $ (increases its opposite; the objective function of $\mathcal{P}1$ \eqref{P1}) without violating the constraints, which fulfills the KKT conditions \cite[Theorem 1]
{Birgin2010}. 
Thus, for a sufficiently high iteration index $m$, the obtained solution $\mathbf{\Psi}_m$ for the first subproblem fulfills the constraints ${g}_i^{(m)}(\mathbf{\Psi}_m,\mathbf{D}_m) <0$ $(\forall i)$, and the Lagrangian multipliers $\{\beta_i\}_{i=1}^{K(L+1)}$ converge to $0$ after sufficient consecutive decrements according to \eqref{betaa}. As a consequence, this yields the complementary slackness condition of a KKT point as well as $\max\{0,{g}_i^{(m)}(\mathbf{\Psi}_m,\mathbf{D}_{m-1}) \} = 0$ $(\forall i)$.
Of note, the case when $\varrho_m$ is unbounded ($\varrho_m\rightarrow \infty$) is not analyzed due to the bounding of $\varrho_m$ by a finite ceiling value $\varrho_{\max}$ as shown in \eqref{varrhoeq}.

Regarding the second subproblem in \eqref{PB2}, for a given $\mathbf{\Psi}_m$ obtained by the AL-based RCG, the convex solver provides a global optimal point $\mathbf{D}^{\star}$ for \eqref{PB2}, which enforces strictly the considered secrecy constraints. Furthermore, since $\varrho_m$ of the BD-RIS subproblem at a higher iteration index converges to a higher value controlling the constraints term, the decrease in the AL becomes equivalent to decreasing the objective function $-f(\mathbf{\Psi},\mathbf{D}) $. As a result, the convex solver provides a solution for the SDP $\mathcal{P}4^{(m)}
$ that either reduces or maintains the AL's objective function ($-f(\mathbf{T})$) with respect to the first subproblem, i.e., $-f(\mathbf{\Psi}_m,\mathbf{D}_{m-1})\geq-f(\mathbf{\Psi}_m,\mathbf{D}_{m})$, which satisfies the KKT conditions. 

From another front, in addition to the monotonicity of $f$ over iterations, one can observe that the latter function is upper-bounded as:
\begin{subequations}
\begin{align}
	f(\mathbf{T})
	&=
	\sum \limits_{l=1}^L  \alpha_l\operatorname{Tr}\left[\overline{\mathbf{H}}_{S T_l} \mathbf{R}_v\right] 
	 \\ & \overset{(a)}{\leq}
	\|\mathbf{R}_v\|_F\sum \limits_{l=1}^L  \alpha_l \|\overline{\mathbf{H}}_{S T_l}^H\|_F \\
	&\overset{(b)}{\leq}  
	\mathrm{Tr}\left[\mathbf{R}_v\right]\sum\limits_{l=1}^L  \alpha_l \|\overline{\mathbf{H}}_{S T_l}^H\|_F
	         \\	
	&\overset{(c)}{\leq}
	P_{\mathrm{S}}\sum\limits_{l=1}^L  \alpha_l \|\overline{\mathbf{H}}_{S T_l}^H\|_F
\end{align}    
\end{subequations}
where \textit{Step (a)} holds by virtue of the Cauchy-Schwarz inequality of two matrices, \textit{Step (b)} is produced due to the fact that for any PSD matrix, $\|\mathbf{R}_v\|_F \leq \operatorname{Tr}\left[\mathbf{R}_v\right]$, and \textit{Step (c)} is because any feasible solution has to satisfy the power budget constraint in \eqref{C2a}. 

Finally, by leveraging the above upper-bound property of $f(\mathbf{\Psi},\mathbf{D})$ (lower-bound of its opposite) along with its monotonicity, the sequence $f\left(\mathbf{T}_{m}\right)$ converges to some finite value $f^{(\infty)}$, i.e., $ f^{(\infty)} \geq  f^* \geq \ldots \geq f\left(\mathbf{T}_{m+1}\right) \geq f\left(\mathbf{T}_{m}\right) \geq \ldots \geq f\left(\mathbf{T}_{1}\right)$, yielding a convergence of the objective function $f\left(\mathbf{T}\right)$ of the original problem in \eqref{P1} to a finite value $f^{\star}$.

    \end{IEEEproof}
\end{theorem}

\begin{algorithm}[h]
\caption{Riemannian Augmented Lagrangian (RAL)-based secure BD-RIS ISAC design.}
\label{algg}
\SetAlgoLined
\KwData{$\mathbf{H}_{SR}$, $\{\mathbf{h}_{RD_k}\}_{k=1}^K$, $\{\zeta_{SD_k}\}_{k=1}^K$, $\{\zeta_{ST_l}\}_{l=1}^L$, $P_{\mathrm{S}}$, $\sigma_{n,D}^2$, $\{\sigma_{D_k}^2\}_{k=1}^{K}$, $\{\sigma_{T_l}^2\}_{l=1}^{L}$, ${G}_{\mathrm{it}}$}
\KwResult{ $\{ \overline{\mathbf{Q}}_k^{\mathrm{(opt)}} \}_{k=1}^K$, $ \mathbf{R}_z ^{\mathrm{(opt)}}$, $\mathbf{\Psi}^{\mathrm{(opt)}}$ }
\Begin{
\string\\ \textcolor{blue}{Initialization} \BlankLine
$m \gets 0$  ,  $\mathbf{R}^{(0)}_z \gets \mathbf{0}_{J_{\mathrm{T}} \times J_{\mathrm{T}}}$ \;
Compute $\{\mathbf{h}_{SD_k}\}_{k=1}^K$ and $\{\mathbf{h}_{ST_l}\}_{l=1}^L$ using \eqref{hsomega} \;
Build a unitary matrix $\mathbf{E}_{SR}$ whose first column is $\mathbf{e}^T(\theta _{SR},\phi_{SR})$
\\$\mathbf{H}_{SD} \gets [ \mathbf{h}_{SD_1}^T ,\ldots, \mathbf{h}_{SD_K}^T  ]^T$ \;
$\mathbf{V}_{\mathrm{ZF}} \gets \mathbf{H}_{SD}^H \left( \mathbf{H}_{SD} \mathbf{H}_{SD}^H \right)^{-1}$ \\
$\{\overline{\mathbf{Q}}_k^{(0)}\}_{k=1}^K\gets \{[ \mathbf{V}_{\mathrm{ZF}} ]_{:,k} \left[ \mathbf{V}_{\mathrm{ZF}} \right]_{:,k}^H\}_{k=1}^K$ \;
$\mathbf{\Psi}_0=\mathbf{E}_{SR}$\; $\mathbf{X}_0=\mathbf{I}_M$\; 
\For{$m\gets1$ \KwTo $G_{\mathrm{it}}$}
   {
   \string\\ \textcolor{blue}{Subproblem 1: BD-RIS scattering matrix} \BlankLine
        \string\\ \textcolor{blue}{a: RCG} \BlankLine
   Compute the Lagrangian and its gradient using \eqref{lagrangian}, \eqref{gradlagrangian}-\eqref{gradconst} \;
   Perform a gradient descent using Algorithms 1 and 2 of \cite{Boumaljournal} to obtain $\mathbf{X}_m$ based on $\mathbf{X}_{m-1}$\;
   $\mathbf{\Psi}_m \gets \mathbf{\Psi}_{m-1} \mathbf{X}_m$ \;
Recompute the channels with $\mathbf{\Psi}_m$ using \eqref{hsomega} \;
 \string\\ \textcolor{blue}{b: Langrangian duals and Penalty factor update} \BlankLine
 \For{$i \gets 1$ \KwTo $K(L+1)$}
{
  Update $\beta_i$ using \eqref{betaa} \;
}  
Update $\varrho_m$ using \eqref{varrhoeq} \;
   \string\\ \textcolor{blue}{Subproblem 2: Beamforming + AN}
   \BlankLine
  Solve the convex SDP $\mathcal{P}4^{(m)}$ in \eqref{PB2} to obtain $\{\overline{\mathbf{Q}}_k^{(m)}\}_{k=1}^K$ and $\mathbf{R}_z^{(m)}$\;
\For{$k \gets 1$ \KwTo $K$}
{
  $\overline{\mathbf{Q}}_k^{(m)} = \mathbf{B} \mathbf{\Lambda} \mathbf{B}^H$ \string\\ \textcolor{blue}{Perform EVD} \BlankLine
  $k^{\ast } \gets \arg \max_{k=1,\ldots ,K}\left[ \mathbf{\Lambda }\right] _{k,k}$
$\overline{\mathbf{Q}}%
_{k}^{(\mathrm{opt},m)} \gets \left[ \mathbf{\Lambda }\right] _{k^{\ast},k^{\ast}}\left[ \mathbf{B}\right]
_{:,k^{\ast }}  $
}
     }
$\{\overline{\mathbf{Q}}_k^{(\mathrm{opt})}\}_{k=1}^K \gets \{\overline{\mathbf{Q}}_k^{(\mathrm{opt},G_{\mathrm{it}})}\}_{k=1}^K$  \;
$\mathbf{R}^{(\mathrm{opt})}_z \gets \mathbf{R}^{(G_{\mathrm{it}})}_z$ \;
$\mathbf{\Psi}^{(\mathrm{opt})} \gets \mathbf{\Psi}_{G_{\mathrm{it}}}$ 
} 
\end{algorithm}



\section{Numerical Evaluation}

This section presents illustrative numerical results for the sensing and
secrecy performance of the studied BD-RIS-aided secure ISAC network. Unless
otherwise mentioned, the adopted system parameter values are given in Table %
\ref{sysparam}. Furthermore, the maximum number of iterations for the
proposed AO-based optimization framework is $G_{\mathrm{it}}=30$, while $%
\sigma _{\Omega }^{2}=\kappa \mathcal{T} B_{w} \mathrm{N}_{\mathrm{F}}$
where $\kappa $ is the Boltzmann constant, $\mathcal{T}=298$ K is the
receiver's temperature, $B_{w}=50$ MHz represents the bandwidth, and $%
\mathrm{N}_{\mathrm{F}}=5$ dB is the noise figure. Additionally, the BD-RIS
and each legitimate user or malicious target are positioned along a given
circle's perimeter whose origin is $S$'s location, while its radius, i.e.,
distance from $S$ to that node, is given in Table \ref{sysparam}. For
instance, the legitimate users are positioned along a circle of a $30$-m
radius from its origin ($S$), while the azimuth angles of the users are
equidistantly set from the interval $[-15^{\circ},5^{\circ}]$. Lastly, we
set the initial BD-RIS scattering matrix as $\mathbf{\Psi}_0=\mathbf{A}_0$,
where $\mathbf{A}_0 \in \mathrm{St}(M, M)$ is a unitary matrix whose first
column is the channel vector $\mathbf{e}(\theta_{SR},\phi_{SR})$, i.e.,
aligned with the $S$-$R$ channel's LoS component.

\begin{table}[h]
\caption{System parameter values}
\label{sysparam}\captionsetup{font=footnotesize} \centering%
\begin{tabular}{|p{2.2cm}|p{1.6cm}||p{1.75cm}|p{1.5cm}|}
\hline\hline
\textbf{Parameter} & \textbf{Value/Range} & \textbf{Parameter} & \textbf{%
Value/Range} \\ \hline\hline
$\lambda$ & $5$ cm & $G_{S,\mathrm{T}}$, $G_{S,\mathrm{R}}$ & $25$ dBi \\ 
\hline
$G_{D_k,\mathrm{R}}$, $G_{T_l,\mathrm{R}}$ $(\forall k,l)$ & $12$ dBi & $%
d_{SR}$ & $22$ m \\ \hline
$d_{SD_k}$ $(\forall k) $ & $30$ m & $d_{ST_l}$ & $[19,25]$ m \\ \hline
$\rho_R,\rho_S$ & $\lambda /2$ & $\phi _{SD_k}$ & $[-15^{\circ},5^{\circ}]$
\\ \hline
$\phi _{ST_l}$ & $-20^{\circ}$ & $\phi _{SR}$ & $-40^{\circ}$ \\ \hline
$\theta _{AB}$ $\left(\forall A,B\right)$ & $0^{\circ}$ & $M_a$, $M_b$ & $8$
\\ \hline
$N_{F}$ & $5$ dB & $P_{\mathrm{S}}$ & $25$ dB \\ \hline
$B$ & $50$ MHz & $J_{\mathrm{T}}$ & $12$ \\ \hline
$J_{\mathrm{R}}$ & $6$ & $\mathcal{T}$ & $298$ K \\ \hline
$K,L$ & $2$ & $\gamma _{T}^{(\max )}$ & $5 $ dB \\ \hline
$\mathcal{K}_{SR},\mathcal{K}_{RD_k}$ & $15$ dB & $\mathcal{K}_{RT_l}$ $%
\left(\forall l \right)$ & $\infty$ (LoS) \\ \hline
$\sigma _{\mathrm{RCS}}^{(l)}$ & $10$ dBsm &  $\varrho_0$ &  $1$ \\ \hline
 $\varrho_{\max}$ &  $10^6$  &  $\beta_i^{(0)}$ ($\forall m$)  & $1$ \\ \hline
 $\epsilon$ & $0.8$ & & 
\end{tabular}%
\end{table}
\begin{figure}[h]
\begin{center}
\vspace*{-.3cm} \includegraphics[scale=.5]{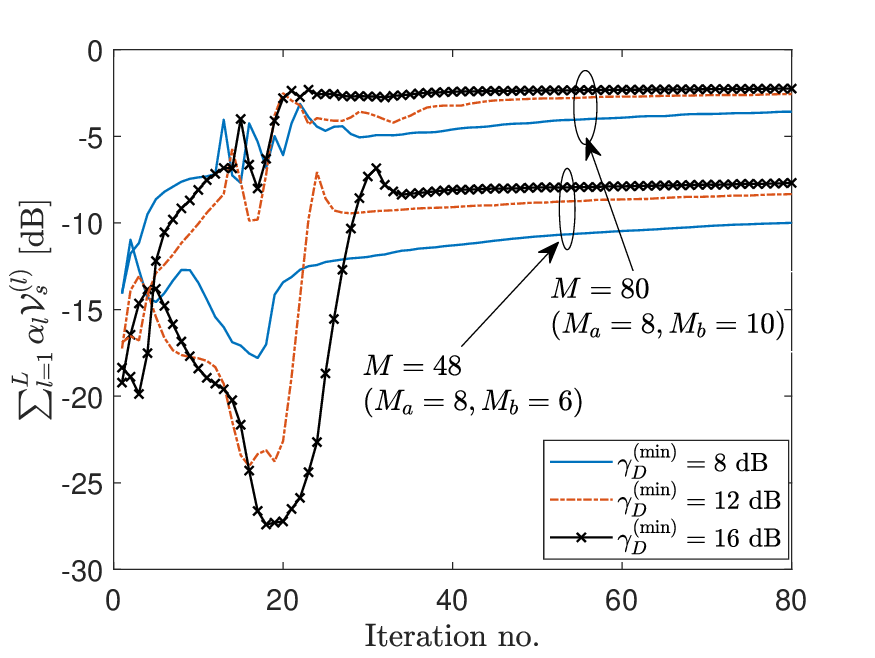} \vspace*{%
-.3cm}
\end{center}
\caption{Evolution of the weighted sum of reflected powers vs. $G_{\mathrm{it%
}}$ for three $\protect\gamma _{D}^{(\min)}$ values.}
\label{fig1}
\end{figure}

In Fig. \ref{fig1}, the objective function of the tackled optimization
problem, given by \eqref{PA}, is shown against the considered AO-based
algorithm's iteration count for different $\gamma_{D}^{(\min)}$ levels. One
can note that the objective function manifests two distinct regimes for the various $M$ and $\gamma_D^{(\min)}$ values, namely a (i) transitory regime and a (ii) convergence saturation one. Note that the former regime characterizes the initial iterations of the proposed iterative AO-based framework, whereby the minimization of the AL in \eqref{lagrangian} is not necessarily achieved through a strict decrease of the objective function, but may instead involve a temporary decrease of the constraint terms, potentially violated at earlier iterations, in order to meet $g_i^{(m)} \rightarrow 0 (\forall i)$. Consequently, the objective function (weighted sum of reflected powers) manifests an unstable fluctuation due to the adaptive constraint violation-dependent penalty factor adjustment. Observe that after a given number of iterations, the objective function stabilizes to a stable value, corresponding to a feasible solution fulfilling the KKT conditions, as demonstrated in the proof of Theorem 1. In particular, around $40$ iterations are sufficient for convergence with $M=80$ REs, whereas around $50$ are needed to achieve a steady value with $M=48$ REs. Such a finding can be justified by the fact that the smaller the BD-RIS, the less likely it is to meet a legitimate SINR constraint. Thus, this yields a longer transitory period in order to balance the AL and fine-tune the weight of the constraint terms. In addition, one can note that the sum reflected power depends on both $M$ and $\gamma_{D}^{(\min)}$, whereas the higher $\gamma_{D}^{(\min)}$ (i.e., stricter system secrecy imposed), the greater the sensing reflected power. In practice, for the considered $M$ values, the increase of $\gamma_{D}^{(\min)}$ from $8$ to $12$ dB yields around $2.5$ dB of sum reflected power enhancement, while raising $\gamma_{D}^{(\min)}$ further from $12$ to $16$ dB yields an extra unit of dB in terms of sensing power. Such a finding demonstrates that the BD-RIS may yield a distinct secrecy-sensing trade-off, compared to the well-known trend in diagonal RIS where imposing stricter secrecy (sensing) requirements degrades the sensing (secrecy) performance \cite{illi2025fdris}.

\begin{figure*}[ht]
\begin{subfigure}[b]{0.33\textwidth}
        \centering
         \includegraphics[scale=.42]{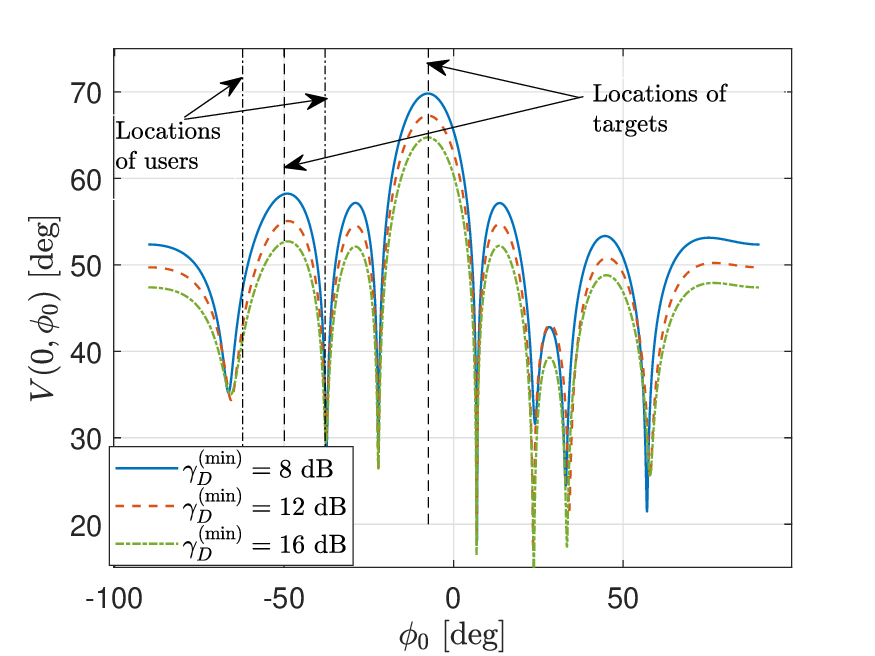}
         \caption{D-RIS.}
         \label{fig2a}
     \end{subfigure}
\begin{subfigure}[b]{0.33\textwidth}
         \centering
         \includegraphics[scale=.42]{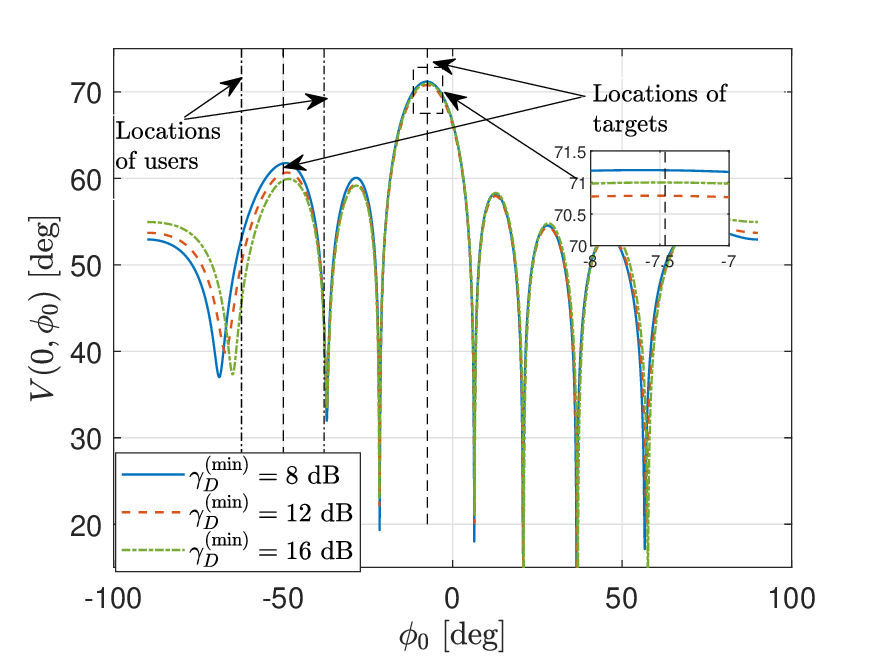}
         \caption{BD-RIS.}
         \label{fig2b}
     \end{subfigure}
\begin{subfigure}[b]{0.33\textwidth}
        \centering
         \includegraphics[scale=.42]{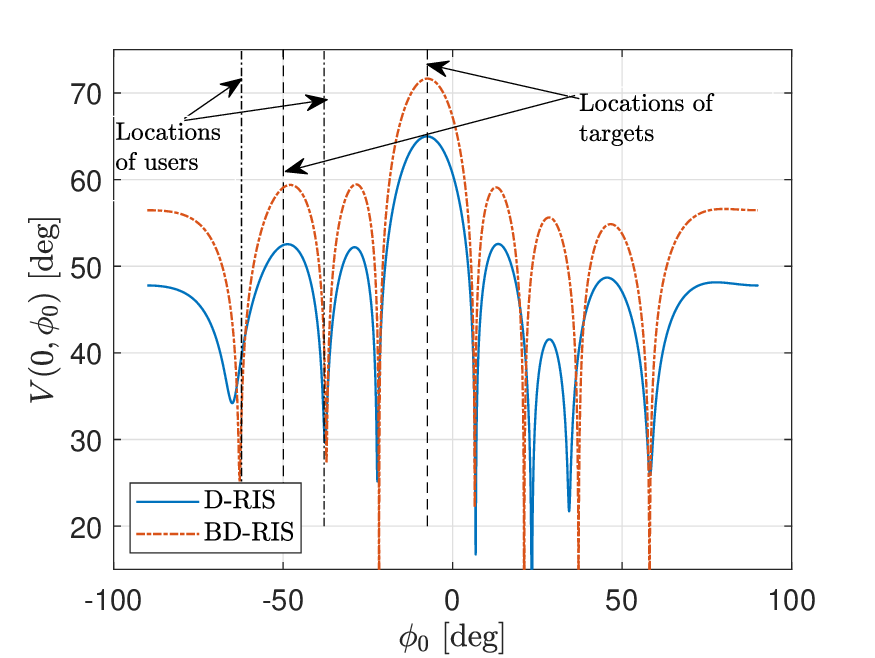}
         \caption{D-RIS vs. BD-RIS for $\gamma_{\mathrm{D}}^{(\min)}=20$ dB.}
         \label{fig2c}
     \end{subfigure}
\caption{The RIS beampattern for the considered BD-RIS-aided ISAC scheme in
comparison with a D-RIS-aided baseline one for $M=64$ REs.}
\label{fig2}
\end{figure*}

In Fig. \ref{fig2}, the considered scheme's BD-RIS beampattern is shown in
terms of the azimuth look direction. Herein, the considered BD-RIS-aided
design is compared against a baseline scheme utilizing a D-RIS. Importantly,
Figs. \ref{fig2a} and \ref{fig2b} evaluate, respectively, the beampattern
gain of a D-RIS and a BD-RIS for three different values of $\gamma_{{D%
}}^{(\min)}$, namely $8$, $12$, and $16$ dB, which correspond to increasing
levels of the network secrecy. It should be highlighted that the D-RIS-based
design was optimized utilizing the considered SDP-based framework in the
secure RIS-aided ISAC scheme of \cite{illi2025fdris}. In addition, Fig. \ref%
{fig2c} compares the considered BD-RIS-aided design with its D-RIS
counterpart in terms of the RIS beampattern profile for $\gamma_{{D%
}}^{(\min)}=20$ dB. Observe in Fig. \ref%
{fig2a} that a conventional D-RIS exhibits main lobes in the directions of
the sensed targets. Notably, the higher $\gamma_{\mathrm{D}}^{(\min)}$ (the
more stricter is the imposed system's secrecy), the lower the RIS
beamforming gain in the direction of the targets. In fact, the broadcasted
ISAC signal by $S$ is reflected by $R$, whereby the AN component is
beamsteered in the directions of the malicious targets to (i) maximize the
sensing reflected power and (ii) prevent the targets from successfully
decoding the legitimate signal, reaching $\{T_l\}_{l=1}^L$ partly via
side lobes. Thus, to fulfill the legitimate SINR constraint in \eqref{C1b},
the D-RIS-based framework reduces the AN power in directions of the
legitimate users to increase the SINRs $\gamma_{D_k}^{(k)}$, which,
consequently, reduces it in the directions of the targets and thus decreases
the sensing power. Such a result demonstrates the existence of a
secrecy-sensing trade-off in traditional D-RIS-based schemes, as concluded
in \cite{illi2025fdris}. In Fig. \ref{fig2b}, despite the similar secrecy-sensing trade-off manifested at the D-RIS, particularly for the first target at $-47^{\circ}$ from $R$'s perspective, the BD-RIS yields a better resilience in terms of the
secrecy-sensing trade-off. In particular, observe that despite the increase
of $\gamma_{{D%
}}^{(\min)}$ from $8$ to $16 $ dB, the beamforming gain only drops by less than $2$ dB, whereas a $5$-dB beamforming gain loss is observed with the D-RIS for the same target. The BD-RIS gains compared to its D-RIS counterpart are more pronounced for the second target located at $-7^{\circ}$ with respect to $R$, where the BD-RIS roughly maintains the same sensing performance, unlike the D-RIS.
A fair comparison between the D-RIS and the BD-RIS is provided in Fig. \ref{fig2c}, which shows an observable $7$-dB gain is observed in favor of the BD-RIS-based scheme.
\begin{figure}[h]
\begin{center}
\vspace*{-.3cm} \includegraphics[scale=.5]{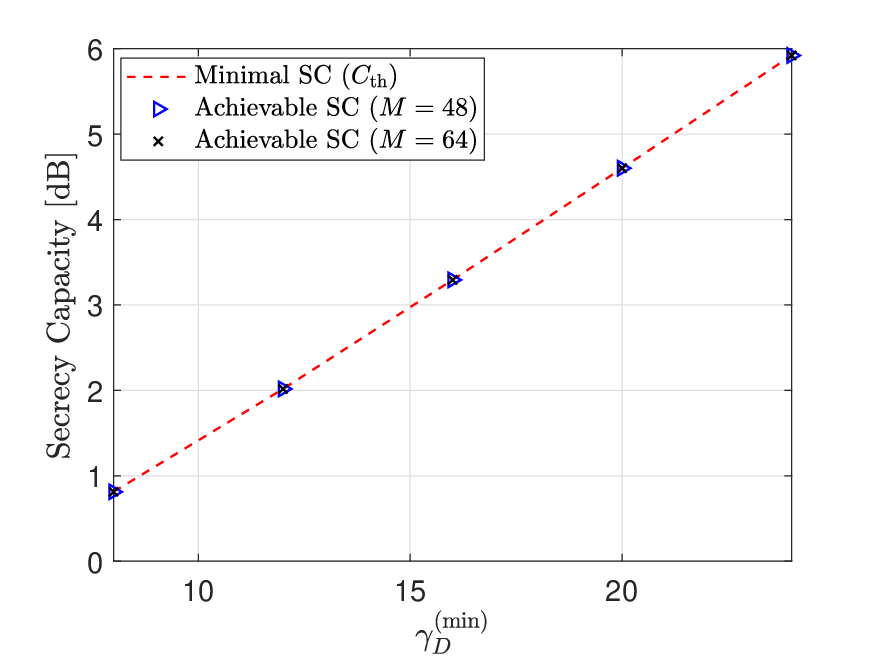} \vspace*{-.3cm}
\end{center}
\caption{Evolution of achievable SC compared of the minimal preset level $C_{%
\mathrm{S}}$ for different $M$ values.}
\label{fig5}
\end{figure}

In Fig. \ref{fig5}, the evolution of the achievable system's SC is shown,
along with the minimal SC threshold level ($C_{\mathrm{th}}$), as a function
of $\gamma_{{D}}^{(\min)}$, and for three different values of $M$. One can
notice that the achievable SC by employing the AO-based optimization
framework is always equal to $C_{\mathrm{th}}$ for the three
considered RIS sizes. Thus, this shows the fulfillment of the secrecy
constraint set in \eqref{C1a}. 
\begin{figure}[h]
\begin{center}
\vspace*{-.3cm} \includegraphics[scale=.5]{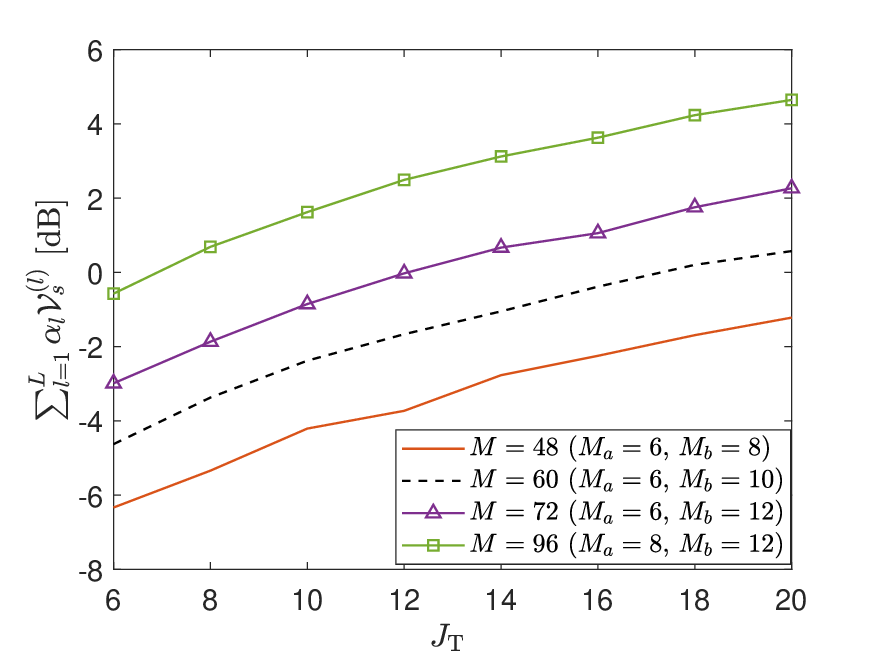} \vspace*{-.3cm}
\end{center}
\caption{Evolution of the weighted sum of reflected powers vs. $J_{\mathcal{T%
}}$ for different $M$ values.}
\label{fig3}
\end{figure}

In Fig. \ref{fig3}, the weighted sum of reflected power is displayed as a
function of the number of transmit antennas ($J_{\mathrm{T}}$) at the BS for
varying values of the BD-RIS size. One can note the increase in the sum of
reflected powers by increasing $J_{\mathrm{T}}$. The adoption of a larger
transmit array results in an improved beamforming gain on the legitimate
signal and the AN one, multiplexed by $S$. This results in a higher
reflected power to the targets by $R$. For instance, a $5$-dB increase of
the sum reflected power is observed when increasing the number of antennas
from $6$ to $20$. Such a beamforming gain can be implicitly noticed from the
ratio $20/6$, which yields: $10 \log_{10}(20/6) \approx 5.22$ dB. The
increase of the BD-RIS size enables additional increase of the sensing
illumination power, where one can notice a consistent $1.75$-, $1.5$-, and $2.5$%
-dB sensing gains, respectively, when increasing $M$ from $48$ to $60$, from $60$ to $72$, and from $72$ to $96$.

\begin{figure}[h]
\begin{center}
\vspace*{-.3cm} \includegraphics[scale=.5]{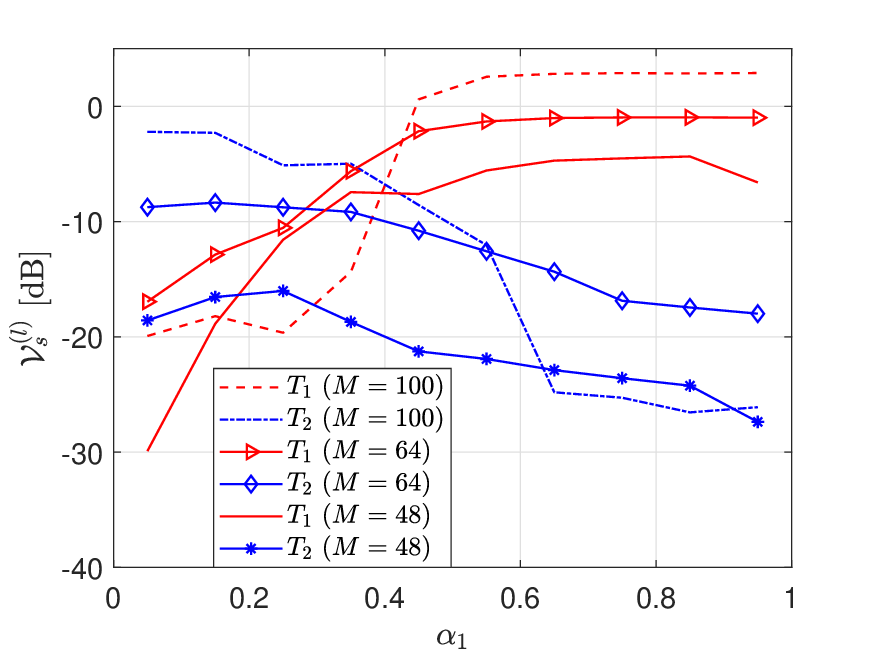} \vspace*{%
-.3cm}
\end{center}
\caption{Per-target reflected power for $L=2$ vs. $\protect\alpha_1$ $(%
\protect\alpha_2=1-\protect\alpha_1)$ for different $M$ values.}
\label{fig4}
\end{figure}

Fig. \ref{fig4} shows the per-target illumination power in terms of the
weight given to $\mathcal{V}_s^{(1)}$, i.e., $\alpha_1$. Notably, in a
two-target case, it follows that $\alpha_2=1-\alpha_1$. Notably, one can
note that the increase of $\alpha_1$ yields a higher reflected power by $T_1$%
, while it decreases for $T_2$. Such an observation means that Algorithm \ref{algg} focuses on maximizing the reflected power with the highest weight to maximize the sum reflected power (objective function). Furthermore, the
intersection point ($\alpha_1^{\star}$) between both targets' power levels,
corresponding to an equilibrium between reflected power at both targets,
depends essentially on the RIS size. Observe that the higher $M$, the higher $%
\alpha_1^{\star}$ is, which approaches $0.4$ at $M=100$ REs. Therefore, larger BD-RIS tend to reach a maximal fairness among targets when setting an equal weight to each target ($\alpha_1=\alpha_2=0.5$).

\section{Conclusion}

This paper proposed a BD-RIS-assisted secure ISAC scheme aiming at
maximizing the sensing performance in the presence of a minimal secrecy
requirement. In particular, an optimization problem was formulated, having
as an objective function a weighted sum of the sensing reflected powers by
the various sensed targets, while constraints on the minimal SC level, the
maximal BS transmit power, and the BD-RIS unitarity property are considered.
An AO-based framework was proposed for iteratively optimizing the network's
BD-RIS scattering matrix, transmit beamforming matrices, and the AN
covariance matrix. Precisely, an RCG-based approach was adopted for
retrieving a suboptimal BD-RIS scattering matrix based on the constructed
AL, while an optimal transmit beamforming and AN covariance matrices
were obtained by virtue of an SDR. The obtained results highlighted the
notable sensing gain offered by the BD-RIS compared with its D-RIS
counterpart, whereby a beampattern gain of at least $7$-dB is observed with $%
M=64$ REs. Also, it was shown that the use of BD-RIS can ensure increased system secrecy without a significant sensing performance degradation, unlike conventional D-RIS.

\bibliographystyle{IEEEtran}
\bibliography{refs}

\end{document}